\documentclass[PRD,preprintnumbers,amsmath,amssymb,floatfix,10pt,prd,twocolumn,superscriptaddress,nofootinbifloatfix,aps,notitlepage,nofootinbib,twocolumn,showpacs,amssymb]{revtex4-1}
\usepackage{latexrelease}
\usepackage{latexsym}
\usepackage{epsfig}
\usepackage{epstopdf}
\usepackage{graphicx}
\usepackage{amssymb}
\usepackage{amsmath}
\usepackage{amsfonts}
\usepackage{subfigure}
\usepackage{dcolumn}
\usepackage{bm}
\usepackage[normalem]{ulem}
\usepackage[dvipsnames]{xcolor}
\usepackage{hyperref}
\usepackage{color}
\usepackage{comment}
\usepackage{float}
\usepackage[utf8]{inputenc}
\usepackage{mathrsfs}
\usepackage{mathtools}
\usepackage{orcidlink}
\usepackage{tikz}
\usepackage{xcolor}
\begin{document}
%------------------------------------------------------------------------------------------------
\title{Regular black holes with Minkowskian cores: energy-condition obstruction and optical screening}
%\title{Regular BHs by gravitational decoupling: a minimal model}
%
\author{Francisco Tello-Ortiz \orcidlink{0000-0002-7104-5746}}
\email{francisco.tello@ufrontera.cl}
\affiliation{Departamento de Ciencias Físicas, Universidad de La Frontera, Casilla 54-D, 4811186 Temuco, Chile.}

\author{Y. Gómez-Leyton}
\email{ygomez@ucn.cl}
\affiliation{Departamento de Física, Universidad Católica del Norte, Av. Angamos 0610, Antofagasta, Chile.
}

\author{Alejandro Rueda
\orcidlink{0009-0005-6302-0038}}
\email{uqarueda@uq.edu.au}
\affiliation{School of Mathematics and Physics, University of Queensland, Brisbane, Queensland 4072, Australia}

\author{Kazuharu Bamba
\orcidlink{0000-0001-9720-8817}}
\email{bamba@sss.fukushima-u.ac.jp}
\affiliation{ Faculty of Symbiotic Systems Science, Fukushima University, Fukushima 960-1296, Japan.}

\author{Manuel González-Espinoza}
\email{manuel.gonzalez@pucv.cl}
\affiliation{
Instituto de Física, Pontificia Universidad Católica de Valparaíso, Casilla 4950, Valparaíso, Chile.}

\begin{abstract}
We construct an asymptotically flat black-hole geometry with a Minkowskian core using gravitational decoupling and analyze its static and rotating extensions. Rather than imposing global energy-condition viability, we show that the vanishing central density required by the Minkowskian core produces a local null-energy-condition obstruction for the present static closure class. The exact mass profile can be written in terms of the regularized incomplete gamma function, which makes transparent both the horizon structure and the exponential localization of the deformation. This localization yields an analytic optical-screening mechanism: on the black-hole branch, deviations of the critical impact parameter from Schwarzschild are controlled by the exponentially suppressed mass deficit at the photon sphere. We also clarify the status of the rotating extension: curvature invariants remain finite on the physical domain $r\geq0$, but this local statement does not by itself establish a globally regular analytic extension through the Kerr disk. These results separate local core regularization from global completeness and explain why strong modifications of the interior can coexist with very small changes in exterior optical observables.
\end{abstract} 
\maketitle
%------------------------------------------------------------------------------------------------
%
\section{Introduction}
Within the framework of general relativity (GR), it has been accepted for decades that gravitational collapse can end in regions where geometric quantities cease to have regular values \cite{Hawking:1973uf,Wald:1984rg}. Singularity theorems show that this situation is not a strange mathematical artifact, but something that occurs almost inevitably when specific hypotheses about energy and causality are met \cite{Penrose:1969pc,Hawking:1973uf}. To prevent these regions from becoming visible, Penrose proposed that the event horizon would act as a natural barrier \cite{Penrose:1969pc}. But, although this proposal mitigates the problem from a global perspective, it does not change the fact that the theory ceases to make sense within the singular region, which remains a controversial issue from a conceptual point of view \cite{Earman:1995fv}.

Over the past few years, various approaches have been proposed to avoid the appearance of singularities in compact solutions. In fact, a considerable range of regular black holes (RBH) models already exists. Most of these proposals resort to some practical matter with unusual properties or to nonlinear electrodynamic formulations, which allow for smoothing the geometric behavior in the innermost region \cite{Salazar:1987ap,Ayon-Beato:1998hmi,Bronnikov:2000vy,Dymnikova:2004zc,Balart:2014cga,Toshmatov:2014nya,Fan:2016hvf,Allahyari:2019jqz}. In general, the singular center is replaced by a finite kernel, which usually approximates a de Sitter geometry quite well \cite{Frolov:2016pav,Hayward:2005gi}. However, this type of regularization brings with it a well-known problem: the presence of internal horizons. These surfaces, known as Cauchy horizons, are not harmless from a dynamic point of view, since they exhibit significant instabilities when reasonable perturbations are considered \cite{Bonanno:2020fgp,Carballo-Rubio:2022kad,Franzin:2022wai,Bonanno:2022jjp,Casadio:2022ndh,OvalleCasadioGiusti2023,Ovalle:2023vvu,Dafermos:2003vim}. 

Interest in these models has increased further thanks to recent advances in observational research. The detection of gravitational waves produced in black hole (BH) mergers by the LIGO–Virgo–KAGRA collaborations \cite{LIGOScientific:2016aoc,LIGOScientific:2017vwq,LIGOScientific:2021djp}, along with the first shadow images obtained by the Event Horizon Telescope (EHT) \cite{EventHorizonTelescope:2019dse,EventHorizonTelescope:2019ggy,EventHorizonTelescope:2019jan,EventHorizonTelescope:2019pgp,EventHorizonTelescope:2019ths,EventHorizonTelescope:2019uob}, have significantly changed how we evaluate the predictions of GR. These results allow us, at least in principle, to test deviations from the Kerr metric in strong-field scenarios and to examine alternative models that incorporate additional degrees of freedom \cite{Johannsen2016,Cunha:2018acu,Psaltis:2018xkc,Vagnozzi:2022moj}. However, several recent studies show that even solutions with very different internal structures can produce shadows virtually indistinguishable from the Kerr shadow \cite{Allahyari:2019jqz,Cunha:2017eoe}, which leaves open the question of how sensitive these global observables really are to deep modifications of space-time.

{In this work, we use gravitational decoupling (GD) to construct a static regular geometry by deforming a Schwarzschild vacuum seed with an additional anisotropic source $\theta_{\mu\nu}$}\footnote{For further applications concerning BH physics, see Ref.~\cite{Ovalle:2018ans,Contreras:2019iwm,Ovalle:2018umz,Ovalle:2020kpd,Fernandes-Silva:2019fez,Casadio:2022ndh,Contreras:2021yxe,Ovalle:2023vvu}}{~\cite{Ovalle:2017fgl,Ovalle:2019qyi}. The GD formalism itself is not claimed as a new ingredient; the model-specific content lies in the closure conditions, localized source profile, resulting Minkowskian-core solution, and its geometric and optical consequences. The rotating spacetime is subsequently obtained by embedding the mass profile derived from the static GD construction into the G\"{u}rses--G\"{u}rsey class.} {A striking feature of the solution is that its interior does not approach a de Sitter kernel but a Minkowskian core, as in the geometries discussed in Refs.~\cite{Simpson:2019mud,Simpson:2018tsi}. The price of this behavior is important: once the effective source is required to vanish at the center while remaining positive immediately outside it, the tangential null energy condition cannot hold in a whole neighborhood of the origin. We therefore treat the energy-condition violation as part of the geometry rather than as an artifact to be hidden. In addition, we show analytically that the deformation is exponentially screened in the photon region, providing a direct explanation for the near-degeneracy of the critical curve with Schwarzschild and, in the rotating exterior, with Kerr.}

{Minkowskian-core constructions obtained from alternative GD closures, their quasinormal-mode and shadow phenomenology, and comparisons between Minkowskian and de~Sitter cores have been studied independently in Refs.~\cite{Hua:2025ovn,LingWu2023,Xiong:2025mink,YangTangXu2024}. These works target closely related questions --- a different choice of~$\mathcal E$ within the same GD scheme~\cite{Hua:2025ovn}, sub-Planckian curvature for Minkowski-core profiles~\cite{LingWu2023}, and an explicit shadow/accretion comparison between Minkowskian and de~Sitter cores~\cite{Xiong:2025mink,YangTangXu2024} --- and we situate the present construction, and in particular the energy-condition obstruction identified below, relative to them rather than treating the Minkowskian core as an isolated novelty.}

Furthermore, a more detailed analysis of the observational characteristics is performed using the ray-tracing algorithm to model the full radiative appearance of a geometrically thin, optically thin accretion disk, including the direct emission and the higher-order photon rings produced by successive disk crossings \cite{Luminet1979,Gralla2019,GLM2020}. {This simulation provides a richer observable than the shadow boundary alone because it encodes redshift and lensing through the transfer function. Doppler effects are included only when a non-static emitter four-velocity is explicitly specified, as in the rotating-disk calculation; they are not attributed to the static-emitter model.}

{The central question addressed here is therefore not merely whether a Minkowskian core can be generated, but which properties follow specifically from the vanishing central density and which are generic consequences of a localized regularization. We show that the two-horizon branch is a standard fold bifurcation of the horizon equation, whereas the Minkowskian core leaves a sharper imprint on the effective matter sector through a local NEC obstruction. Conversely, the exterior optical response is exponentially screened because the mass profile has already saturated to its ADM value well inside the photon region.}

{The paper is organized as follows. Section~\ref{sec2} summarizes the GD framework and the corresponding field equations. Section~\ref{sec3} constructs and analyzes the static regular geometry, including its energy conditions, horizon structure, and inner causal region. Section~\ref{sec4} introduces the G\"{u}rses--G\"{u}rsey rotating extension, determines its black-hole domain and extremal locus, and studies its critical curves. Section~\ref{sec5} examines the ray-traced appearance of thin, optically thin emission models in the static and rotating geometries. Our conclusions are collected in Sec.~\ref{sec6}.}

\section{Gravitational Decoupling}
\label{sec2}

{In this section we summarize the gravitational-decoupling (GD) framework used below. The formalism itself is not a new ingredient of the present work: we follow the extended geometric-deformation scheme developed in Refs.~\cite{Ovalle:2017fgl,Ovalle:2019qyi}. Its role is to organize the Einstein equations into a known seed sector and an additional effective source. The model-specific input is introduced only in Sec.~\ref{sec3}, where the Schwarzschild seed, the closure conditions, and the localized density profile are specified. Related extensions to axially symmetric configurations can be found in Ref.~\cite{Contreras:2021yxe}.}

\subsection{Field equations and matter sector}

We begin with the Einstein--Hilbert action containing two independent matter 
Lagrangians,
\begin{equation}
S=\int \left[ \frac{R}{2\kappa} + \mathcal{L}_{M} + \mathcal{L}_{\Theta} \right] 
\sqrt{-g}\, d^4x ,
\label{action}
\end{equation}
{where $\mathcal{L}_{\rm M}$ denotes the seed matter sector and $\mathcal{L}_{\Theta}$ the additional sector. No microscopic interpretation of $\mathcal{L}_{\Theta}$ is required for the construction below; we treat $\theta_{\mu\nu}$ as an effective anisotropic source. The corresponding energy--momentum tensors are}
\begin{align}
T_{\mu\nu} &= -\frac{2}{\sqrt{-g}} \frac{\delta (\sqrt{-g}\,\mathcal{L}_{\rm{M}})}
{\delta g^{\mu\nu}}
= -2\frac{\delta\mathcal{L}_{\rm{M}}}{\delta g^{\mu\nu}} + g_{\mu\nu}\mathcal{L}_{\rm{M}}, 
\label{TM}\\[1ex]
\theta_{\mu\nu} &= -\frac{2}{\sqrt{-g}} \frac{\delta (\sqrt{-g}\,\mathcal{L}_{\Theta})}
{\delta g^{\mu\nu}}
= -2\frac{\delta\mathcal{L}_{\Theta}}{\delta g^{\mu\nu}} + g_{\mu\nu}\mathcal{L}_{\Theta}.
\label{Ttheta}
\end{align}
The complete system then satisfies the Einstein equations
\begin{equation}
G_{\mu\nu} = \kappa\, \tilde T_{\mu\nu},
\qquad
\tilde T_{\mu\nu} = T_{\mu\nu} + \theta_{\mu\nu},
\label{Einstein}
\end{equation}
with the total energy--momentum tensor obeying the conservation law
\begin{equation}
\nabla^{\mu}\tilde T_{\mu\nu} = 0,
\label{conservation}
\end{equation}
as implied by the Bianchi identities.

\subsection{Spherically symmetric geometry}

We restrict to static and spherically symmetric configurations described by the metric
\begin{equation}
ds^{2}
= -e^{\nu(r)} dt^{2}
+ e^{\lambda(r)} dr^{2}
+ r^{2} d\Omega^{2},
\label{metric_general}
\end{equation}
with $d\Omega^{2}=d\theta^{2}+\sin^{2}\theta\, d\phi^{2}$.  
In this background, Eqs.~\eqref{Einstein} yield
\begin{align}
\kappa\,\tilde T^{0}{}_{0} &=
-\frac{1}{r^{2}}
+ e^{-\lambda}\left(\frac{1}{r^{2}} - \frac{\lambda'}{r} \right),
\label{Einstein00}
\\
\kappa\,\tilde T^{1}{}_{1} &=
-\frac{1}{r^{2}}
+ e^{-\lambda}\left(\frac{1}{r^{2}} + \frac{\nu'}{r} \right),
\label{Einstein11}
\\
\kappa\,\tilde T^{2}{}_{2} &=
\frac{e^{-\lambda}}{4}
\left[
2\nu'' + \nu'^{2} - \lambda'\nu'
+ \frac{2}{r}(\nu' - \lambda')
\right],
\label{Einstein22}
\end{align}
with $\tilde T^{3}{}_{3}=\tilde T^{2}{}_{2}$ by symmetry.  
These equations allow the standard identification of the effective energy density, radial 
pressure, and tangential pressure
\begin{equation}
\tilde\epsilon = -\tilde T^{0}{}_{0}, 
\qquad
\tilde p_{r} = \tilde T^{1}{}_{1}, 
\qquad
\tilde p_{\theta} = \tilde T^{2}{}_{2}.
\end{equation}

By decomposing these contributions into the seed sector and the additional $\theta_{\mu\nu}$ 
source, one has
\begin{align}
\tilde\epsilon &= \epsilon + \mathcal{E} ,
\\
\tilde p_r &= p_r + \mathcal{P}_r ,
\\
\tilde p_\theta &= p_\theta + \mathcal{P}_\theta ,
\end{align}
where $T^{\mu}_{\nu}=\mathrm{diag}(-\epsilon,p_r,p_\theta,p_\theta)$ and
$\theta^{\mu}_{\nu}=\mathrm{diag}(-\mathcal{E},\mathcal{P}_r,\mathcal{P}_\theta,\mathcal{P}_\theta)$.

\subsection{Decoupling the system}

{The starting point of GD is a known geometry sourced by $T_{\mu\nu}$. Let $(\xi,\mu)$ denote the metric functions of this seed solution,}
\begin{equation}
ds^{2}=-e^{\xi(r)}dt^{2}+e^{\mu(r)}dr^{2}+r^{2}d\Omega^{2},
\label{seed_metric}
\end{equation}
with
\begin{equation}
e^{-\mu(r)} = 1 - \frac{2m(r)}{r},
\qquad
m(r) = \frac{\kappa}{2}\int_{0}^{r} x^{2}\epsilon(x)\, dx ,
\label{seed_mass}
\end{equation}
the Misner--Sharp mass.

The GD scheme introduces two metric deformations,
\begin{align}
\nu &= \xi + \alpha\, g(r), 
\label{temporal_def}
\\
e^{-\lambda} &= e^{-\mu} + \alpha\, f(r),
\label{radial_def}
\end{align}
{where $\alpha$ controls the additional sector and $(f,g)$ encode the radial and temporal geometric deformations associated with $\theta_{\mu\nu}$. These relations define a deformation of the seed solution and are not coordinate transformations. No perturbative expansion in $\alpha$ is assumed.}

{Substituting Eqs.~\eqref{temporal_def}--\eqref{radial_def} into Eqs.~\eqref{Einstein00}--\eqref{Einstein22} separates the field equations into a seed system and a second set associated with the additional source:}

\begin{itemize}
\item[(i)] The original Einstein equations for the seed matter $T_{\mu\nu}$,
\begin{align}
\kappa\,\epsilon &= \frac{1}{r^{2}} - e^{-\mu}\left(\frac{1}{r^{2}} - \frac{\mu'}{r}\right),
\label{seed_eq1}
\\
\kappa\,p_r &= -\frac{1}{r^{2}} + e^{-\mu}\left(\frac{1}{r^{2}} + \frac{\xi'}{r}\right),
\label{seed_eq2}
\\
\kappa\,p_\theta &= \frac{e^{-\mu}}{4}
\left[
2\xi'' + \xi'^{2} - \mu'\xi' + \frac{2}{r}(\xi' - \mu')
\right].
\label{seed_eq3}
\end{align}

\item[(ii)] {A second set determining the response associated with the additional tensor $\theta_{\mu\nu}$,}
\begin{align}
\kappa\,\mathcal{E} &= -\alpha\left(\frac{f}{r^{2}} + \frac{f'}{r}\right),
\label{theta1}
\\
\kappa\,\mathcal{P}_r - \alpha Z_{1} &= \alpha f\left( \frac{1}{r^{2}} + \frac{\nu'}{r} \right),
\label{theta2}
\\
\kappa\,\mathcal{P}_\theta - \alpha Z_{2} &=
\frac{\alpha f}{4}
\left(
2\nu'' + \nu'^2 + \frac{2\nu'}{r}
\right)
- \frac{\alpha f'}{4}
\left(
\nu' + \frac{2}{r}
\right),
\label{theta3}
\end{align}
where
\begin{align}
Z_{1} &= \frac{e^{-\mu} g'}{r},
\\
4Z_{2} &= e^{-\mu}
\left(
2g'' + \alpha g'^2 + \frac{2g'}{r}
+ 2g' \xi' - \mu' g'
\right).
\end{align}
\end{itemize}

\noindent
{The seed geometry is recovered continuously for $f=g=0$. Notice that this sector decomposition does not in general imply two independently conserved matter systems when the temporal deformation is nonzero.}

Finally, substituting the metric deformations into the conservation law 
\eqref{conservation} yields
\begin{equation}
\nabla^{\mu} T_{\mu\nu} 
= -\frac{\alpha g'}{2}\,(\epsilon+p_{r})\,\delta^{1}{}_{\nu},
\qquad
\nabla^{\mu}\theta_{\mu\nu}=-\nabla^{\mu} T_{\mu\nu},
\label{exchange}
\end{equation}
{Thus a nontrivial temporal deformation permits an exchange between the seed and additional sectors while the total source remains conserved. The exchange vanishes when $g'=0$ or when the seed satisfies $\epsilon+p_r=0$, including the vacuum seed used below. This statement should not be confused with the closure condition subsequently imposed on the additional source. In particular, GD by itself does not select the density profile, equation of state, or regular core of the present model; those are model-specific ingredients introduced in Sec.~\ref{sec3}.}

\section{Regular BHs}\label{sec3}
\label{sec3}
{We now specialize the GD framework to the construction studied in this work. We choose the Schwarzschild vacuum as the seed geometry,} 
\begin{equation}
	e^\xi
	=
	e^{-\mu}
	=
	1-\frac{2M}{r}
	\label{yscxx}
\end{equation}  
{which solves the seed equations with $T_{\mu\nu}=0$. The additional source $\theta_{\mu\nu}$ is therefore responsible for the deformation away from Schwarzschild. At this stage the GD equations do not determine a unique geometry: the five functions $f$, $g$, ${\cal E}$, ${\cal P}_r$, and ${\cal P}_\theta$ require two independent closure conditions. The construction below consists in specifying these conditions and then selecting a localized density profile whose regularity condition produces a Minkowskian core. Accordingly, the novelty claimed here is not the GD decomposition itself, but the resulting regular solution and its geometric and optical consequences.}
\subsection{Horizon structure}

{As the first closure condition, we restrict the final static geometry to the Schwarzschild gauge,}
\begin{equation}
\label{yconstxx}
e^{\nu}=e^{-\lambda}
\end{equation}
which yields
\begin{equation}
{\cal P}_{r}
	=
	-{\cal E}
	\label{yschwconxx1}
\end{equation}
and therefore, for the $\theta_{\mu\nu}$ sector we have
\begin{eqnarray}
	\label{yxxx}
	{\cal P}_r'=
	\frac{2}{r}\left({\cal P}_\theta-{\cal P}_r\right)
\end{eqnarray}
Next, by using the condition~\eqref{yconstxx} and the seed Schwarzschild solution~\ref{yscxx}, we obtain
\begin{equation}
\label{yxx}
\alpha\,f
=
\left(1-\frac{2M}{r}\right)
\left[e^{\alpha\,g(r)}-1\right]
\end{equation}
Hence the line element~\eqref{metric_general} becomes
\begin{eqnarray}
\label{yhairyxx}
ds^{2}
&=&
-\left(1-\frac{2M}{r}\right)h(r)dt^{2}
+\left(1-\frac{2M}{r}\right)^{-1}\frac{dr^2}{h(r)}
\nonumber
\\
&&
+r^{2}d\Omega^2,
\end{eqnarray}
with
\begin{equation}
\label{yhxx}
h=e^{\alpha\,g},
\end{equation}
where the function $g$ remains unknown.
\subsection{Energy conditions and the Minkowskian-core obstruction}
{Energy conditions provide a useful diagnostic of the effective source~\cite{Martin-Moruno:2017exc}. We begin with the weak energy condition (WEC), not as an assumption that will be enforced globally, but in order to determine precisely where it is satisfied or violated:}
\begin{eqnarray}
\label{yweakxx11}
&&
{\cal E}
\geq
0,
\\
&&
{\cal E}+{\cal P}_r
\geq
0,
\label{yweakxx12}
\\
&&
{\cal E}+{\cal P}_\theta
\geq
0.
\label{yweakxx22}
\end{eqnarray}
Equation ~\eqref{yweakxx12} holds as a consequence of Eq. ~\eqref{yschwconxx1}, while the
condition~\eqref{yweakxx11} becomes
\begin{equation}
	\label{yG1xx}
\kappa\,r^2{\cal E}
=
-\left(r-2M\right)h'-h+1
\geq
0,
\end{equation}
and \ref{yweakxx22} is written as
\begin{eqnarray}
	\label{yG2xx}
	2\left({\cal E}+{\cal P}_\theta\right)
	=
	-r{\cal E}'
	\geq
	0,
\end{eqnarray} 
where $h=h(r)$ is defined in Eq.~\eqref{yhxx}.
We see that Eq.~\eqref{yG1xx} is a first-order linear differential inequality for the deformation $h$, where the case ${\cal E}=0$ leads to the seed Schwarzschild solution in Eq.~\eqref{yscxx}.
{Equation~\eqref{yG2xx} is particularly restrictive for a Minkowskian core. A positive density that vanishes at the center must increase somewhere in an arbitrarily small right neighborhood of $r=0$; wherever ${\cal E}'>0$, the tangential NEC combination ${\cal E}+{\cal P}_\theta=-r{\cal E}'/2$ is negative. Thus, within the present class with ${\cal P}_r=-{\cal E}$, a nontrivial positive Minkowskian-core density entails a local NEC violation. We choose the regular localized profile}
\begin{equation}
\label{yGxx}
\kappa\,{\cal E}
=
\frac{\alpha\,r^2}{\ell^4}\,e^{-r/\ell}.
\end{equation}
{where $\ell$ is a constant with dimensions of length. Notice that ${\cal E}(0)=0$ and ${\cal E}'(r)>0$ for $0<r<2\ell$, so the WEC/NEC cannot be globally satisfied by this profile. At this stage $\alpha$ parametrizes the source in the decoupling construction; after the regularity condition is imposed below, its physical role is more naturally that of an inverse core-size parameter.}
\subsection{Regular spacetime}
Putting together the Eqs. \eqref{yGxx} and \eqref{yG1xx}, we find for the limiting case, the following expression for the function $h(r)$
\begin{eqnarray}
	\label{ystronxx}
	&&h
	=
	\frac{r-c_1}{r-2M}+
	\frac{\alpha\,e^{-r/\ell}}{\ell^3(r-2M)}
	\bigg(24\ell^4+24\ell^3r+12\ell^2\,r^2\nonumber\\
	&&+4\ell\,r^3+r^4\bigg),\nonumber\\
\end{eqnarray}
where the constant $c_1$ is a length. Hence, the asymptotically flat metric functions is given as
\begin{equation}
\begin{split}
	\label{yweakxx}
	e^{\nu}
	=
	e^{-\lambda}
	&=
	1-\frac{c_1}{r} 
	+	\frac{\alpha\,e^{-r/\ell}}{\ell^3\,r}
\bigg(24\ell^4+24\ell^3r+12\ell^2\,r^2\\&+4\ell\,r^3+r^4\bigg).
\end{split}
\end{equation}
Notice that the ADM mass
is given by ${\cal M}=c_1/2$, and for $r\sim 0$ we have
\begin{equation}
	\label{zero}
	e^{\nu}
	=
	e^{-\lambda}
	\simeq
	1+\frac{c_1-24\,\alpha\,\ell}{r}+{\cal O}(r/\ell)^4
	\ .
\end{equation}
Therefore, it is evident that the absence of singularity requires $c_1=24\,\alpha\,\ell$, which leads to
\begin{equation}
	\label{reg}
	12\alpha\,\ell
	=
	{\cal M}.
\end{equation}

{So, after imposing the regularity condition in Eq.~\eqref{reg}, the metric function behaves as
\begin{equation}
e^{\nu}(r)=e^{-\lambda}(r)=1+\mathcal{O}(r^{4}),
\qquad r\rightarrow 0.
\end{equation}
Therefore, the spacetime approaches Minkowski space at the center. The same condition was found in Ref.~ \cite{Simpson:2019mud,Simpson:2018tsi}. This behavior contrasts with the majority of regular BH constructions, where the central geometry is instead de Sitter-like, typically characterized by
\begin{equation}
e^{\nu}(r)=1-\frac{\Lambda_{\rm eff}}{3}r^{2}+\mathcal{O}(r^{4}).
\end{equation}
Such a de Sitter core appears in many regular BH models supported by nonlinear electrodynamics or effective matter sources, including the Bardeen-type and Hayward geometries and their extensions in Ref.~\cite{Bardeen1968,Hayward2006,BalartVagenas2014}. 
The absence of the quadratic term in the present model implies that the regularization mechanism does not rely on an effective vacuum energy density at the origin. Instead, the geometry is supported by a matter distribution whose energy density vanishes smoothly at the center. 
This feature places the present solution closer to regular BH geometries with a Minkowskian core, such as the Minkowski-core models discussed in Refs.~\cite{OvalleCasadioGiusti2023,Simpson:2019mud,Simpson:2018tsi}.}
{It is important, however, to distinguish the present construction from those precedents. Minkowskian cores and exponentially suppressed mass profiles are not by themselves new. In the present model the profile is instead fixed by the GD closure together with the localized density in Eq.~\eqref{yGxx}, and regularity converts it into the incomplete-gamma mass function \eqref{massgamma}. The results emphasized below are consequently not the mere existence of a flat core, but the local NEC obstruction implied by ${\cal P}_r=-{\cal E}$ and ${\cal E}(0)=0$, and the analytic exterior-screening law that follows from the exact mass deficit $Q(5,r/\ell)$. This separation between construction, core physics, and exterior optical response is the point on which our comparison with earlier Minkowski-core geometries is based.}

Finally, we write the metric in Eq.~\ref{yweakxx} in terms of the ADM mass given by Eq.~\ref{reg} as 
\begin{eqnarray}
	\label{yweakxx2}
	&&	e^{\nu}
	=
	e^{-\lambda}
	=
	1-\frac{2\,{\cal M}}{r}\nonumber\\
	&&	+	\frac{\alpha^4\,e^{-12\alpha\,r/{\cal M}}}{{\cal M}\,r}
	\bigg(1728\frac{r^{4}}{\mathcal{M}^{2}}+2\frac{\mathcal{M}^{2}}{\alpha^{4}}+72\frac{\mathcal{M}r}{\alpha^{3}} \nonumber\\ && +144\frac{r^{2}}
    {\alpha^{2}}+576\frac{r^{3}}{\mathcal{M}\alpha}\bigg).
\end{eqnarray}

The mass function now reads,

\begin{equation}\label{mass1}
\begin{split}
    \tilde{m}(r)=\mathcal{M}-\frac{\alpha^4\,e^{-12\alpha\,r/{\cal M}}}{2{\cal M}}
	\bigg(1728\frac{r^{4}}{\mathcal{M}^{2}}+2\frac{\mathcal{M}^{2}}{\alpha^{4}}+72\frac{\mathcal{M}r}{\alpha^{3}} \\ +144\frac{r^{2}}
    {\alpha^{2}}+576\frac{r^{3}}{\mathcal{M}\alpha}\bigg),
    \end{split}
\end{equation}
{Introducing
\begin{equation}
x\equiv\frac{r}{\ell}=\frac{12\alpha r}{\mathcal M},
\end{equation}
the same result takes the compact form
\begin{equation}
\tilde m(r)=\mathcal M P(5,x)
=\mathcal M\left[1-e^{-x}\sum_{k=0}^{4}\frac{x^k}{k!}\right],
\label{massgamma}
\end{equation}
where $P(s,x)=\gamma(s,x)/\Gamma(s)$ is the regularized lower incomplete gamma function. Consequently,
\begin{equation}
F(r)=1-\frac{2\mathcal M}{r}P(5,x),
\qquad
\mathcal M-\tilde m(r)=\mathcal M Q(5,x),
\label{Fgamma}
\end{equation}
with $Q=1-P$. This representation makes explicit that the source is exponentially localized and that the mass function saturates rapidly to the ADM mass. After regularity is imposed, $\ell/\mathcal M=(12\alpha)^{-1}$; hence $\alpha\to+\infty$ is the Schwarzschild limit, whereas $\alpha\to0$ sends the core scale to infinity and yields Minkowski spacetime.}

\begin{figure}[H]
    \centering     \includegraphics[scale=0.6]{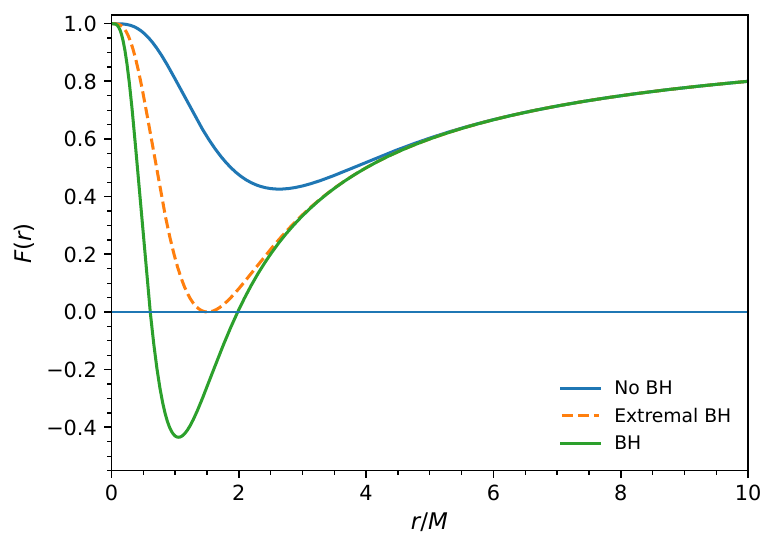}
        \caption{The inverse radial metric potential versus the radial coordinate for $\mathcal{M}=1$ and $\alpha=0.2$ for a horizonless configuration (red line), $\alpha=\alpha_{E}$ for the extremal BH (blue line) and $\alpha=0.5$ for the two horizon BH spacetime (black line).}  
    \label{fig1}
\end{figure}

In Fig.~\ref{fig1}, the trend of the inverse radial metric potential is depicted, where for $r\to0$ the metric function $e^{-\lambda}\to1$.

\begin{figure}[H]
    \centering     \includegraphics[scale=0.7]{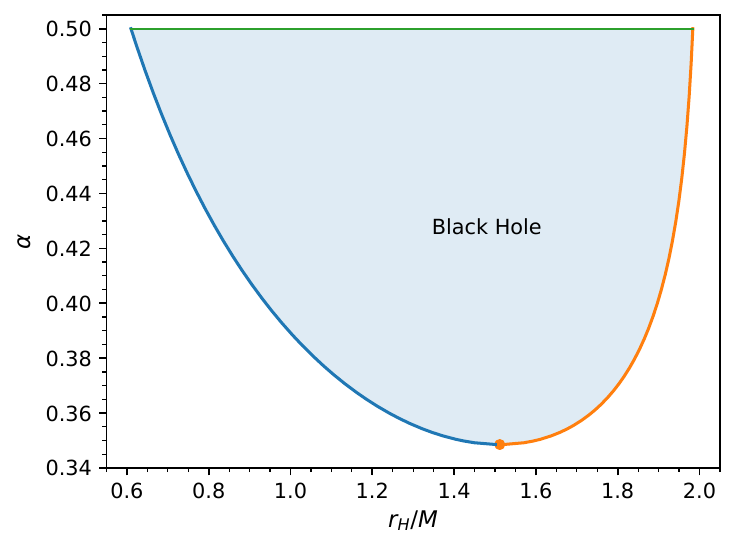}
        \caption{The BH region showing the admissible values for the parameter $\alpha$. }  
    \label{fig2}
\end{figure}

Now, we analyze the causal structure of the spacetime obtained. As we filled the Schwarzschild vacuum spacetime with the $\theta_{\mu\nu}$ source, it is expected that this additional source substantially modifies the number of horizons of the seed model. Of course, the solution of the horizon equation 
\begin{equation}\label{eh}
    e^{-\lambda(r)}\bigg|_{r=r_{H}}=0,
\end{equation}
is quite involved and requires numerical solution. In this case, the roots of Eq.~\eqref{eh} depend on $\cal{M}$ and $\alpha$. In Fig.~\ref{fig1} the inverse radial metric potential is shown versus the radial coordinate. As can be observed, the BH presents an inner and outer horizons (black line), an extremal case (blue line) and no horizon (red line). Of course, these situations occur for particular values of the deformation parameter $\alpha$. The extremal condition requires \cite{Mayo:1996mv,Hod:2022mys}
\begin{equation}\label{extremalcondition}
    e^{-\lambda(r)}\bigg{|}_{r=r_{H}}=0, \quad \left[\frac{de^{-\lambda(r)}
    }{dr}\right]\bigg{|}_{r=r_{H}}=0.
\end{equation}
Solving numerically the above conditions for $\{r_{H},\alpha\}$ considering $\mathcal{M}=1$, one obtains
\begin{equation}
    r_{H}\approx1.512,\quad \alpha_{E}\approx0.348.
\end{equation}
{
The extremal configuration obtained from Eq.~\eqref{extremalcondition} plays the role of a critical point separating two qualitatively different branches of the solution. Let us define
\begin{equation}
F(r,\alpha)\equiv e^{-\lambda,\alpha}(r),
\end{equation}
so that horizons are determined by the equation
\begin{equation}
F(r,\alpha)=0.
\end{equation}
At the extremal point $(r_E,\alpha_E)$ one has
\begin{equation}
F(r_E,\alpha_E)=0,
\qquad
\partial_r F(r_E,\alpha_E)=0.
\label{ExtremalConditionsAgain}
\end{equation}
Expanding $F(r,\alpha)$ around the extremal configuration,
\begin{equation}
F(r,\alpha)\simeq 
\frac{1}{2}\left(\partial_r^2 F\right)_E (r-r_E)^2
+\left(\partial_\alpha F\right)_E (\alpha-\alpha_E)
+\cdots,
\label{TaylorExtremal}
\end{equation}
where the subscript $E$ indicates evaluation at $(r_E,\alpha_E)$, one immediately finds that the nearby horizons satisfy
\begin{equation}
r_{\pm}-r_E
\simeq
\pm
\sqrt{
-\,2
\frac{(\partial_\alpha F)_E}{(\partial_r^2 F)_E}
(\alpha-\alpha_E)
}.
\label{HorizonSplitting}
\end{equation}
This expression makes the branch structure transparent: for one sign of $(\alpha-\alpha_E)$ no real roots exist, corresponding to the no-horizon branch, while for the opposite sign two distinct real roots appear, corresponding to the inner and outer horizons of the BH branch. The extremal solution is therefore the bifurcation point at which the two-horizon geometry merges into a double root.} In Fig.~\ref{fig2} the BH region is shown. It is observed that for those values of $\alpha<\alpha_{E}$ there is not a BH solution.  

Next, we check the regularity of some curvature invariants. The first one is the Ricci scalar given by
\begin{equation}
R=\frac{A{e}^{-\frac{12 r \alpha}{\mathcal{M}}} r^2 \alpha^5(2r\alpha-\mathcal{M})}{\mathcal{M}^5},
\end{equation}
and Ricci squared reads

\begin{equation}
R_{\mu\nu}R^{\mu\nu}=\frac{B e^{-\frac{24 r \alpha}{\mathcal{M}}} r^4 \alpha^{10}\left(5 \mathcal{M}^2-24 \mathcal{M} r \alpha+36 r^2 \alpha^2\right)}{\mathcal{M}^{10}},
\end{equation}
{where the coefficients are $A=-124416$ and $B=859963392$. Direct differentiation of $F=1-2\tilde m/r$ reproduces both expressions. Both invariants vanish at the center. The Kretschmann scalar is finite as well; using $\ell=\mathcal M/(12\alpha)$ its near-center expansion can be written equivalently as
\begin{equation}
K=\frac{212}{25}\frac{\alpha^2 r^4}{\ell^8}+\mathcal O(r^5)
=\frac{91156119552}{25}\frac{\alpha^{10}r^4}{\mathcal M^8}
+\mathcal O(r^5),
\end{equation}
so the static geometry is curvature regular and approaches a flat core.}

{To close the static analysis, we evaluate the energy conditions without assuming their sign. From Eq.~\eqref{yGxx},
\begin{equation}
\kappa {\cal E}'=\frac{\alpha r}{\ell^4}e^{-r/\ell}\left(2-\frac{r}{\ell}\right),
\end{equation}
and therefore
\begin{equation}
{\cal E}+{\cal P}_\theta
=-\frac{r}{2}{\cal E}'
=\frac{41472\,e^{-12\alpha r/\mathcal M}r^2\alpha^5(6r\alpha-\mathcal M)}{2\kappa\mathcal M^5}.
\label{eq51}
\end{equation}
Thus ${\cal E}+{\cal P}_\theta<0$ for $0<r<2\ell=\mathcal M/(6\alpha)$, vanishes at $r=2\ell$, and becomes positive outside that radius. Since ${\cal E}+{\cal P}_r=0$, the NEC and WEC are violated only through the tangential sector in the inner core. This is not an accidental feature of the chosen exponential profile: for any regular positive density with ${\cal E}(0)=0$ and ${\cal P}_r=-{\cal E}$, the density must increase somewhere close to the origin and Eq.~\eqref{yG2xx} then forces a tangential NEC violation there.

The dominant energy condition (DEC) provides an additional restriction. Using
\begin{equation}
\frac{{\cal P}_\theta}{{\cal E}}=\frac{x}{2}-2,
\qquad x=\frac{r}{\ell},
\end{equation}
the condition $|{\cal P}_\theta|\leq {\cal E}$ is satisfied only for
\begin{equation}
2\leq x\leq6.
\end{equation}
Hence the effective source should not be interpreted as a globally conventional material fluid. In particular, formal quantities such as $d{\cal P}_i/d{\cal E}$ are not automatically physical propagation speeds in the absence of a microscopic barotropic matter model; we therefore do not use them as causality or stability criteria.}

\subsection{Inner horizon and causal structure}

{The black-hole branch possesses an outer event horizon $r_+$ and an inner Cauchy horizon $r_-$. Their existence follows directly from the global shape of $F(r)$. Regularity gives $F(0)=1$, while asymptotic flatness gives $F(\infty)=1$. Therefore, if the unique finite-radius minimum of $F$ crosses zero, two positive roots are generated; at the critical value $\alpha=\alpha_E$ they merge into a double root. The local expansion in Eq.~\eqref{HorizonSplitting} is consequently the standard square-root splitting associated with a fold (saddle-node) bifurcation. We do not attribute this bifurcation itself to a new causal mechanism.}

{What distinguishes the present model from de Sitter-core regular black holes is instead the geometry and matter content inside the inner horizon. Here $F(r)=1+\mathcal O(r^4)$ and the effective density vanishes at $r=0$, whereas a de Sitter core has $F(r)=1-\Lambda_{\rm eff}r^2/3+\cdots$ and a finite vacuum-energy density. Thus the Cauchy horizon should not be interpreted as the boundary of an effective de Sitter region. Regularity of the center and multiplicity of horizons are logically distinct: the former fixes the small-$r$ expansion, while the latter is controlled by whether the finite-radius minimum of $F$ lies above, at, or below zero.}

{The dynamical stability of the Cauchy horizon is a separate question. The present analysis is purely stationary and does not establish suppression of mass inflation or of the usual blueshift instability. Any such claim requires a perturbative calculation and is therefore left open.}

\section{Axially symmetric case}
\label{sec4}
As spherically symmetric and static BHs are just toy models to get an idea about how the gravitational interaction is in the strong field regime. To capture a more realistic situation and provide a deeper insight about more realistic BH models, we construct here the rotating version of the model \eqref{yweakxx2}. To do so, we follow the methodology presented
in Ref.~\cite{Contreras:2021yxe}
(see also Refs.~\cite{Burinskii:2001bq,Dymnikova:2006wn,Smailagic:2010nv,Bambi:2013ufa,Azreg-Ainou:2014nra,Dymnikova:2016nlb}).
We consider the general Kerr-Schild metric in Boyer-Lindquist coordinates, namely, the G\"{u}rses-G\"{u}rsey metric~\cite{Gurses:1975vu}
\begin{eqnarray}
	\label{kerrex}
	ds^{2}
	&=&
	-\left[1-\frac{2\,r\,{\tilde{m}}(r)}{{\rho}^2}\right]
	dt^{2}
	-
	\frac{4\, {a}\, r\,{\tilde{m}}(r)\, \sin^{2}\theta}{{\rho}^{2}}
	\,dt\,d\phi
	\nonumber
	\\
	&&
	+
	\frac{{\rho}^{2}}{{\Delta}}\,dr^{2}
	+
	{\rho}^{2}\,d\theta^{2}
	+
	\frac{{\Sigma}\, \sin^{2}\theta}{{\rho}^{2}}\,d\phi^{2}
	\ ,
\end{eqnarray}
\begin{figure}[H]
    \centering     \includegraphics[width=0.5\textwidth]{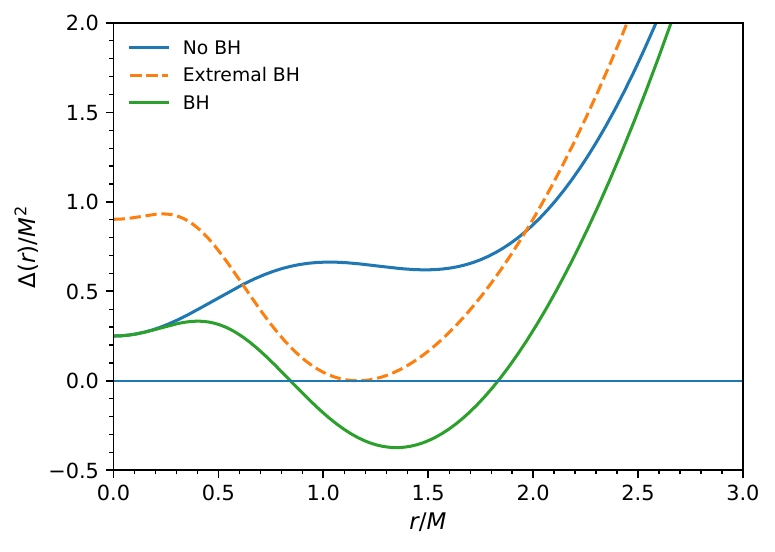}
        \caption{The $\Delta(r)$ function versus the radial coordinate $r$ for $\mathcal M=1$. The black curve corresponds to a black hole with $\{a,\alpha\}=\{0.5,0.5\}$, the blue curve to the extremal black hole with $\{a,\alpha\}=\{0.95,0.71192\}$, and the red curve to the horizonless configuration $\{a,\alpha\}=\{0.5,0.3\}$.}  
    \label{fig3}
\end{figure}

where 
\begin{eqnarray}
	%2\,\tilde{m}&=&r-\left[r-2\,m(r)\right]h(r)\ ,
	%	\\
	{\rho}^2
	&=&
	r^2+{a}^{2}\cos^{2}\theta,
	\label{f0}
	\\
	{\Delta}
	& = &
	r^2-2\,r\,{\tilde{m}}(r)
	+{a}^{2},
	\label{f2}
	\\
	{\Sigma}
	& = &
	\left(r^{2}+{a}^{2}\right)^{2}
	-{\Delta}\, a^2\sin^{2}\theta,
	\label{f3}\\
	{a} & = & {J}/{\cal M}.
\end{eqnarray}
Here, $\tilde m$ is the mass function of the  metric~\eqref{yweakxx2},
${J}$ is the angular momentum and ${\cal M}$ the total mass of the system.
Obviously, Eq.~\eqref{kerrex} reproduces the well-known Kerr solution when $\tilde m=\mathcal{M}$. Rotating extensions of regular BHs have been extensively investigated in the literature, often constructed through Newman–Janis–type algorithms or alternative regularization procedures. Representative examples include the rotating regular BH solutions proposed in \cite{BambiModesto2013} and the more general construction developed in \cite{AzregAinou2014}. 

{The rotating geometry considered here belongs to this general class of Kerr-like spacetimes. More precisely, we embed the mass profile $\tilde m(r)$ derived from the static GD construction into the G\"{u}rses--G\"{u}rsey metric. We therefore do not claim that the rotating line element follows directly from a new GD decoupling of the stationary axisymmetric field equations; the link with GD is inherited through the static mass profile.}

The horizon structure is determined by solving the equation 
\begin{equation}\label{eq57}
    \Delta(r_{H})=r^{2}_{H}-2r_{H}\tilde{m}(r_{H})+a^{2}=0, 
\end{equation}
where the function $\tilde m(r)$ plays the role of a radially varying mass profile which replaces the constant mass parameter of the Kerr solution. This modification regularizes the geometry near the center while preserving the Kerr asymptotic structure at large distances. Similar mechanisms appear in several rotating regular BH constructions discussed in the literature \cite{BambiModesto2013,AzregAinou2014}.

\begin{figure}[H]
    \centering     \includegraphics[width=0.45\textwidth]{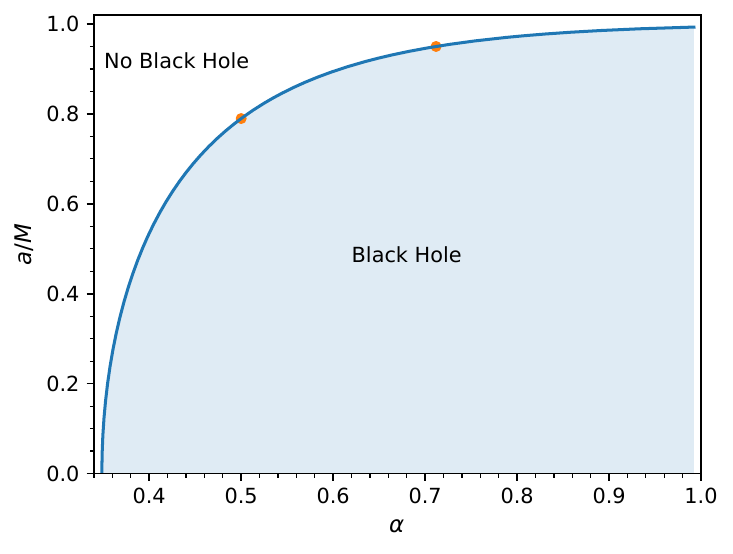}
        \caption{{Rotating parameter space in the $(a,\alpha)$ plane for $\mathcal M=1$. The blue curve is the extremal locus obtained by simultaneously solving $\Delta(r_E;a,\alpha)=0$ and $\partial_r\Delta(r_E;a,\alpha)=0$. It separates the black-hole branch, for which $\Delta=0$ has horizons, from the horizonless branch.}}  
    \label{fig4}
\end{figure}
{The horizon boundary in the $(a,\alpha)$ plane must be determined from the simultaneous equations
\begin{equation}
\Delta(r_E;a,\alpha)=0,
\qquad
\partial_r\Delta(r_E;a,\alpha)=0.
\label{rot_ext}
\end{equation}
For example, setting $\mathcal M=1$ and $a=0.95$ gives $r_E\simeq1.16411$ and $\alpha_{\rm ext}\simeq0.71192$. Conversely, at $\alpha=0.7$ the extremal spin is $a_{\rm ext}\simeq0.94585$ with $r_E\simeq1.17199$. Hence the parameter set $(a,\alpha)=(0.95,0.7)$ previously used as ``extremal'' is actually horizonless. In the Schwarzschild/Kerr limit $\alpha\to\infty$, the extremal spin approaches $a/\mathcal M\to1$; it should not be identified with a finite value of $\alpha$. Figure~\ref{fig4} is to be understood as the numerical locus defined by Eq.~\eqref{rot_ext}.}

\begin{figure*}
    \centering         \includegraphics[width=0.98\textwidth]{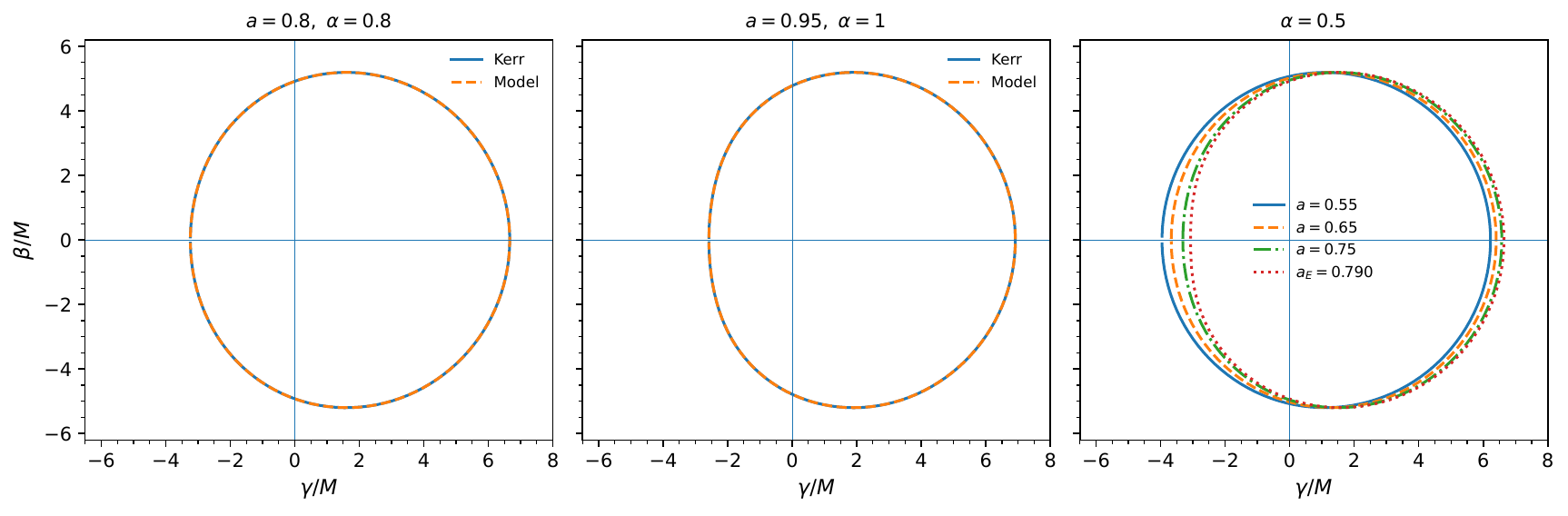}
        \caption{{Critical curves obtained from the unstable spherical-photon-orbit conditions \eqref{radialpotential}--\eqref{celestial} for $\mathcal M=1$ and an equatorial observer, $\iota=90^\circ$. \textbf{Left}: present geometry (dashed red) and Kerr (black) for $(a,\alpha)=(0.8,0.8)$. \textbf{Middle}: the same comparison for $(a,\alpha)=(0.95,1)$. Both model configurations lie on the two-horizon black-hole branch. \textbf{Right}: critical curves of the present geometry at fixed $\alpha=0.5$ for $a=0.55$, $0.65$, $0.75$, and the extremal value $a_{\rm ext}\simeq0.78967$. The last curve is the double-root limit of $\Delta$ at this value of $\alpha$.}}
    \label{fig5}
\end{figure*}

{For completeness, we state explicitly how the critical curves in Fig.~\ref{fig5} are obtained. The Hamilton--Jacobi equation for null geodesics in the G\"{u}rses--G\"{u}rsey geometry retains the Kerr-type separated radial potential
\begin{equation}
{\cal R}(r)=\left(r^2+a^2-a\xi\right)^2
-\Delta(r)\left[(\xi-a)^2+\eta\right],
\label{radialpotential}
\end{equation}
where $\xi=L_z/E$ and $\eta={\cal Q}/E^2$. Unstable spherical photon orbits satisfy ${\cal R}=0=\partial_r{\cal R}$, which gives
\begin{align}
\xi(r)&=\frac{r^2+a^2-4r\Delta/\Delta'}{a},\\
\eta(r)&=\frac{16r^2\Delta}{(\Delta')^2}-[\xi(r)-a]^2.
\label{shadowimpact}
\end{align}
For an observer at infinity at inclination $\iota$, the celestial coordinates are
\begin{equation}
\gamma=-\frac{\xi}{\sin\iota},\qquad
\beta=\pm\sqrt{\eta+a^2\cos^2\iota-\xi^2\cot^2\iota}.
\label{celestial}
\end{equation}
Thus the equatorial choice used in Fig.~\ref{fig5} reduces to $\gamma=-\xi$ and $\beta=\pm\sqrt{\eta}$. We have independently checked the horizon status of every parameter set displayed in that figure: $(a,\alpha)=(0.8,0.8)$ and $(0.95,1)$ both possess two horizons, while at fixed $\alpha=0.5$ the sequences $a=0.55$, $0.65$, and $0.75$ lie on the black-hole branch and $a_{\rm ext}\simeq0.78967$ is the limiting double-root configuration. The model--Kerr curves in the first two panels are therefore compared within the admissible black-hole domain.}

{We deliberately do not attach diameter, area, or circularity percentages to Fig.~\ref{fig5} unless a specific estimator for each quantity is defined and evaluated from the final numerical curve. The robust statement supported by the figure is the near-overlap of the critical curves for the displayed parameter choices. This behavior is consistent with the same localization mechanism identified analytically in the static sector: once the mass function has nearly saturated in the photon region, the exterior null dynamics becomes weakly sensitive to the detailed core profile. In the rotating case this is a qualitative screening interpretation rather than the static first-order formula \eqref{bcshift}, which should not be transferred unchanged to Kerr-like photon orbits. {Were a circularity deviation to be reported in a future version, the RMS estimator of Ref.~\cite{Bambi:2019tjh}, $\Delta C=\bar R^{-1}\{(2\pi)^{-1}\int_0^{2\pi}d\phi\,[R(\phi)-\bar R]^2\}^{1/2}$, rather than a mean-absolute-deviation surrogate, should be used.}}

{For the G\"{u}rses--G\"{u}rsey class the Ricci scalar can be written directly in terms of the mass profile as
\begin{equation}
R=\frac{2\left(2\tilde m'(r)+r\tilde m''(r)\right)}{\rho^2}
=-\frac{124416\,e^{-12 r\alpha/\mathcal M}r^4\alpha^5(2r\alpha-\mathcal M)}
{\mathcal M^5\rho^2}.
\label{Rrot}
\end{equation}
The second equality follows by differentiating Eq.~\eqref{massgamma} and provides a direct algebraic check of the coefficient and radial scaling.}
{The regularity statement can be made considerably sharper without displaying prohibitively long expressions after substituting the exponential mass profile. For a G\"{u}rses--G\"{u}rsey metric with an arbitrary mass function $\tilde m(r)$, define
\begin{align}
\lambda_1&=\frac{2a^2\cos^2\theta\,\tilde m'(r)+r\rho^2\tilde m''(r)}{\rho^4},\\
\lambda_2&=\frac{2r^2\tilde m'(r)}{\rho^4}.
\label{riceigen}
\end{align}
The Ricci tensor has the two doubly degenerate eigenvalues $\lambda_1$ and $\lambda_2$, and therefore
\begin{equation}
R_{\mu\nu}R^{\mu\nu}=2\left(\lambda_1^2+\lambda_2^2\right).
\label{Ricci2rot}
\end{equation}
The Weyl sector is encoded in the type-D scalar
\begin{equation}
\Psi_2=
-\frac{6\varrho\,\tilde m
-2(r+\varrho)\bar\varrho\,\tilde m'
+r\bar\varrho^{\,2}\tilde m''}
{6\varrho\,\bar\varrho^{\,3}},
\qquad
\varrho=r+ia\cos\theta,
\label{Psi2rot}
\end{equation}
so that the Kretschmann scalar can be evaluated compactly as
\begin{equation}
K\equiv R_{\alpha\beta\gamma\delta}R^{\alpha\beta\gamma\delta}
=48\,{\rm Re}\!\left(\Psi_2^2\right)
+2R_{\mu\nu}R^{\mu\nu}-\frac{1}{3}R^2.
\label{Krot}
\end{equation}
Equations~\eqref{Rrot}, \eqref{Ricci2rot}, and \eqref{Krot} provide an exact representation of the three standard quadratic curvature diagnostics for the present rotating solution.}

{For our mass function the small-$r$ behavior is stronger than the usual $O(r^3)$ regular-core condition:
\begin{equation}
\tilde m(r)=c_5 r^5+\mathcal O(r^6),
\qquad
c_5=\frac{\mathcal M}{120\ell^5}
=\frac{10368}{5}\frac{\alpha^5}{\mathcal M^4}.
\label{mrotseries}
\end{equation}
Hence $\tilde m(0)=\tilde m'(0)=\tilde m''(0)=0$. To test the potentially dangerous Kerr ring without selecting a privileged direction, introduce
\begin{equation}
\zeta\equiv\frac{a\cos\theta}{r},
\end{equation}
and approach $(r,\theta)=(0,\pi/2)$ while keeping an arbitrary finite limiting value of $\zeta$. Direct substitution of Eq.~\eqref{mrotseries} gives
\begin{align}
R&=\frac{60c_5}{1+\zeta^2}\,r^2+\mathcal O(r^3),\\
R_{\mu\nu}R^{\mu\nu}
&=
\frac{200c_5^2\left(9\zeta^4+12\zeta^2+5\right)}
{\left(1+\zeta^2\right)^4}\,r^4+\mathcal O(r^5),\\
K&=
\frac{16c_5^2\left(225\zeta^8+402\zeta^6+470\zeta^4
+250\zeta^2+53\right)}
{\left(1+\zeta^2\right)^6}\,r^4+\mathcal O(r^5).
\label{rotinvariantseries}
\end{align}
All three invariants therefore vanish at the ring for every finite directional ratio $\zeta$. Approaches with $|\zeta|\to\infty$, corresponding to reaching the $r=0$ disk away from the equatorial ring or approaching the ring more slowly in angle than in radius, are also finite and give a vanishing limit. Thus the apparent factors of $\rho^{-2}$ and $\rho^{-4}$ do not generate a scalar-curvature singularity. As a consistency check, the equatorial limit $\zeta=0$ gives
\begin{equation}
K=848\,c_5^2 r^4+\mathcal O(r^5)
=\frac{212}{25}\frac{\alpha^2r^4}{\ell^8}+\mathcal O(r^5),
\end{equation}
which coincides with the static near-center coefficient.}

{More generally, the present profile satisfies the known necessary and sufficient local conditions for finiteness of all second-order scalar polynomial curvature invariants of this Kerr-like class, namely $\tilde m(0)=\tilde m'(0)=\tilde m''(0)=0$. The calculation above is stronger for the three standard diagnostics used here: because the leading mass term is $O(r^5)$, $R$, $R_{\mu\nu}R^{\mu\nu}$, and $K$ actually vanish at the would-be Kerr ring. This establishes absence of a scalar-curvature singularity on the physical $r\geq0$ branch. It does not, however, establish geodesic completeness or a smooth global extension through the $r=0$ disk. Indeed, Eq.~\eqref{massgamma} contains both odd and even powers beyond leading order, and its formal continuation to $r<0$ develops exponentially growing terms as $r\to-\infty${, a pathology of exactly the kind identified for the rotating Hayward metric by Zhou and Modesto~\cite{ZhouModesto2023}, who show that analytic continuation through the Kerr-type disk is generically obstructed unless $\tilde m(r)$ is an odd function of $r$ (which it is not here); see also Torres and Fayos~\cite{TorresFayos2017} for the regularity conditions required of $\tilde m(r)$ in this Kerr--Schild class}. We therefore keep the global-extension question separate from the local curvature-regularity result.

{We have likewise not extended the energy-condition analysis of Sec.~\ref{sec3} off the equatorial plane for the rotating metric. On general grounds the NEC-violating region found in the static, equatorial sector is expected to enlarge away from the equator for a Kerr--Schild rotating profile built on the present mass function, as shown for this class of regular rotating black holes by Neves and Saa~\cite{NevesSaa2014}, who prove that rotation unavoidably enlarges the weak-energy-condition violation relative to the static seed. A direct off-equatorial computation of $\tilde\epsilon+\tilde p_\perp$ for the present $\tilde m(r)$, analogous to Eq.~\eqref{yG2xx}, is left for a revised version.}}

\begin{figure*}
    \centering
    \includegraphics[width=0.32\linewidth]{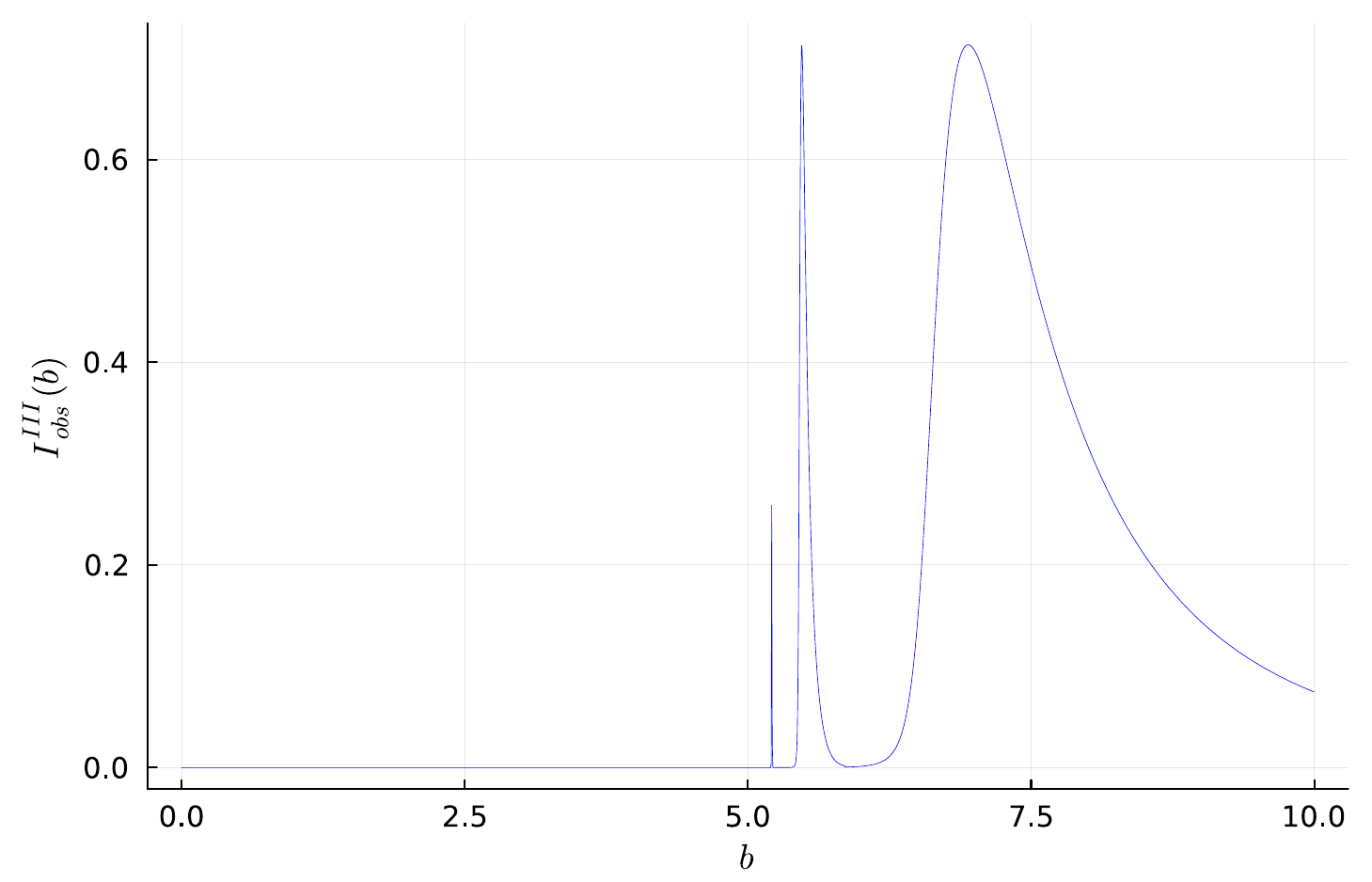}\,
    \includegraphics[width=0.32\linewidth]{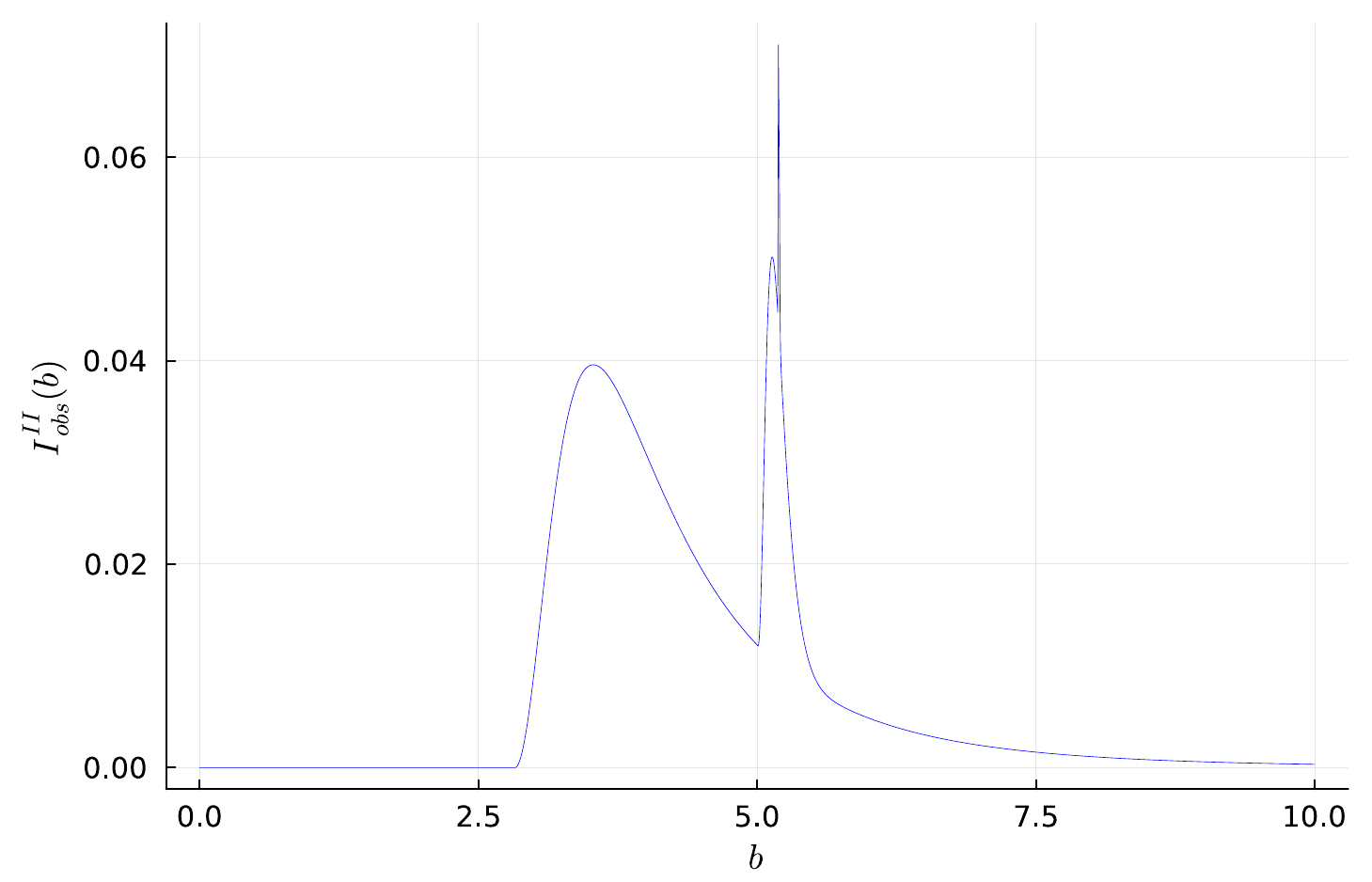}\,
    \includegraphics[width=0.32\linewidth]{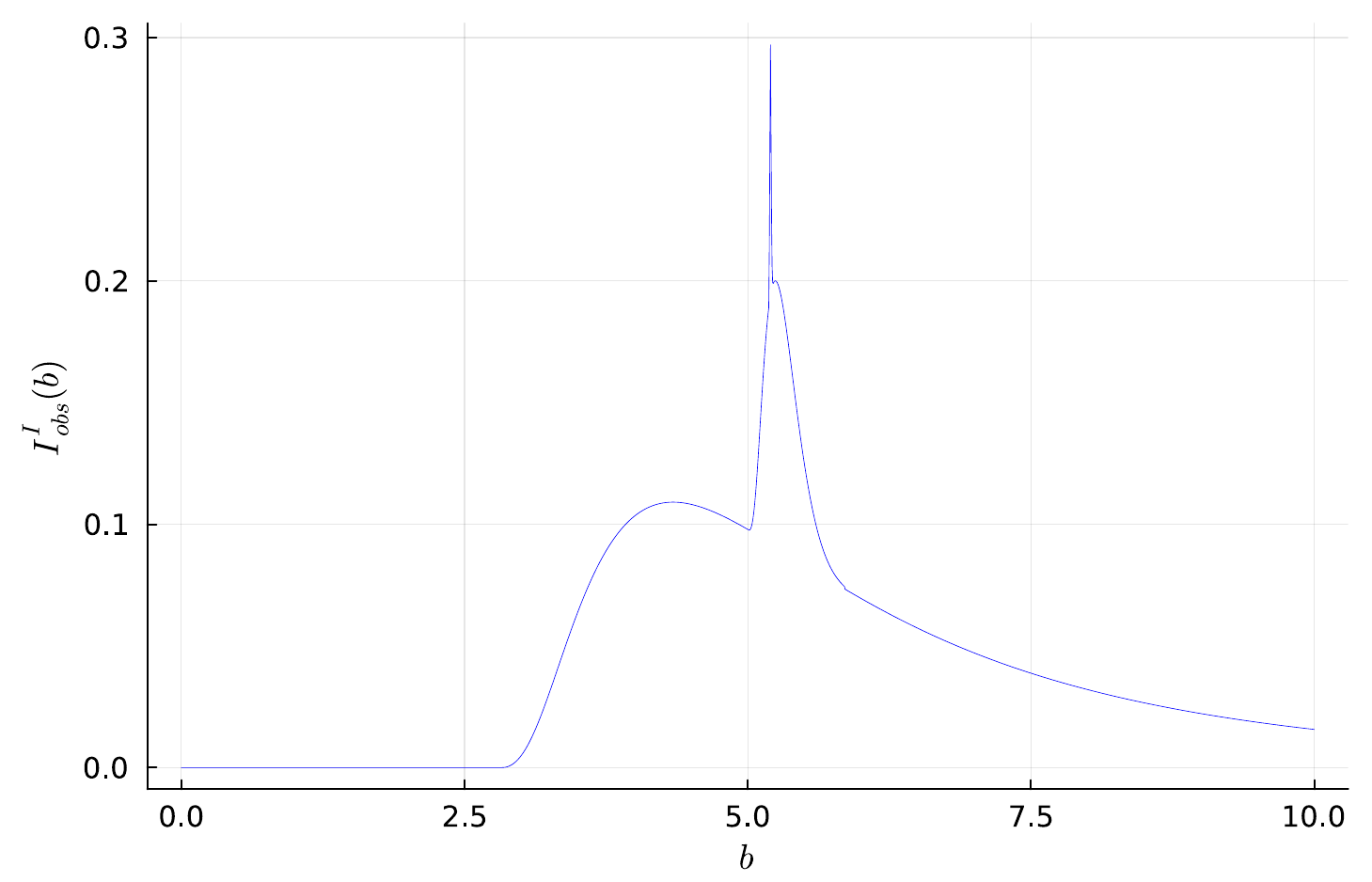}\,
    \includegraphics[width=0.32\linewidth]{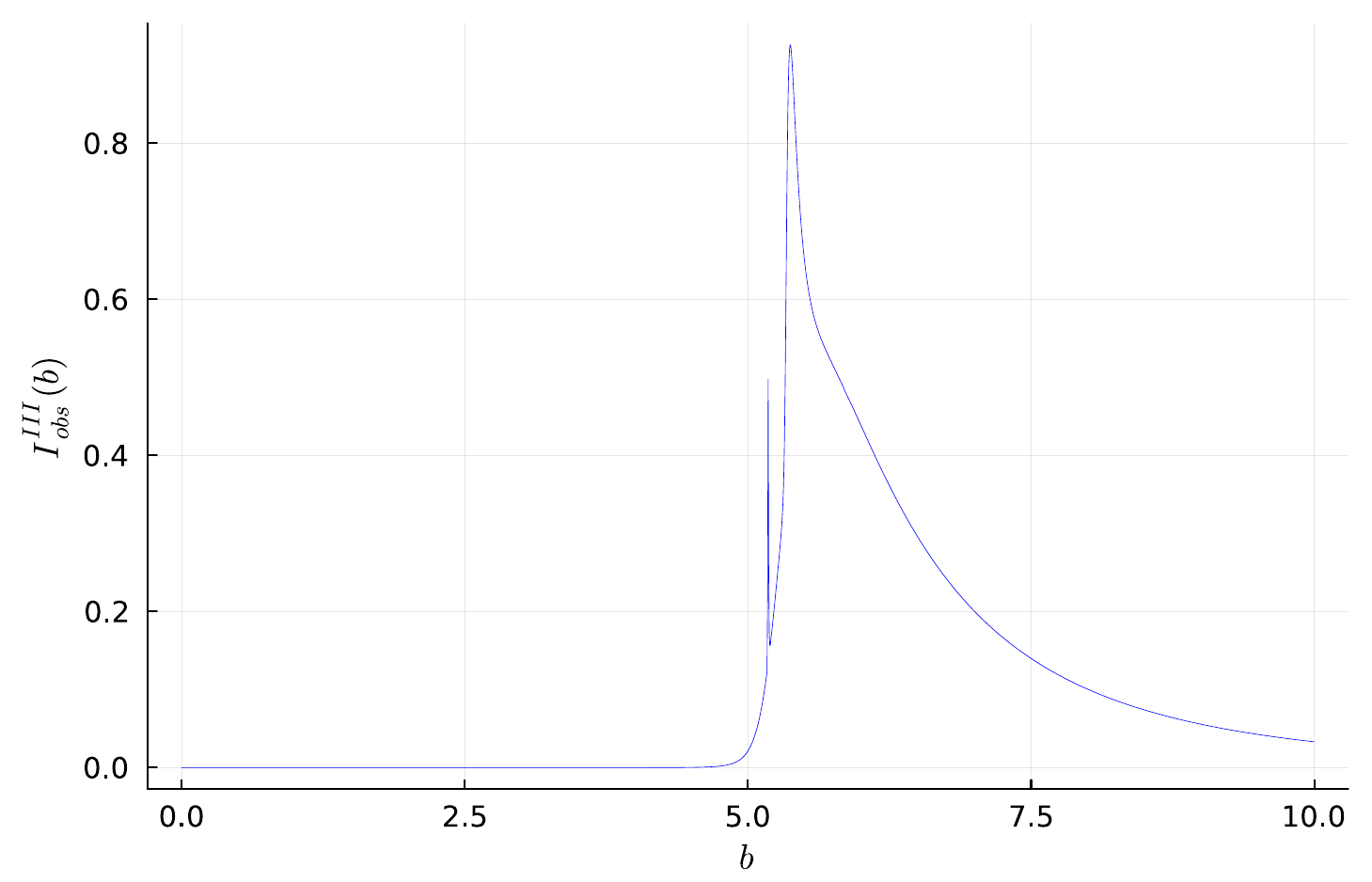}\,    \includegraphics[width=0.32\linewidth]{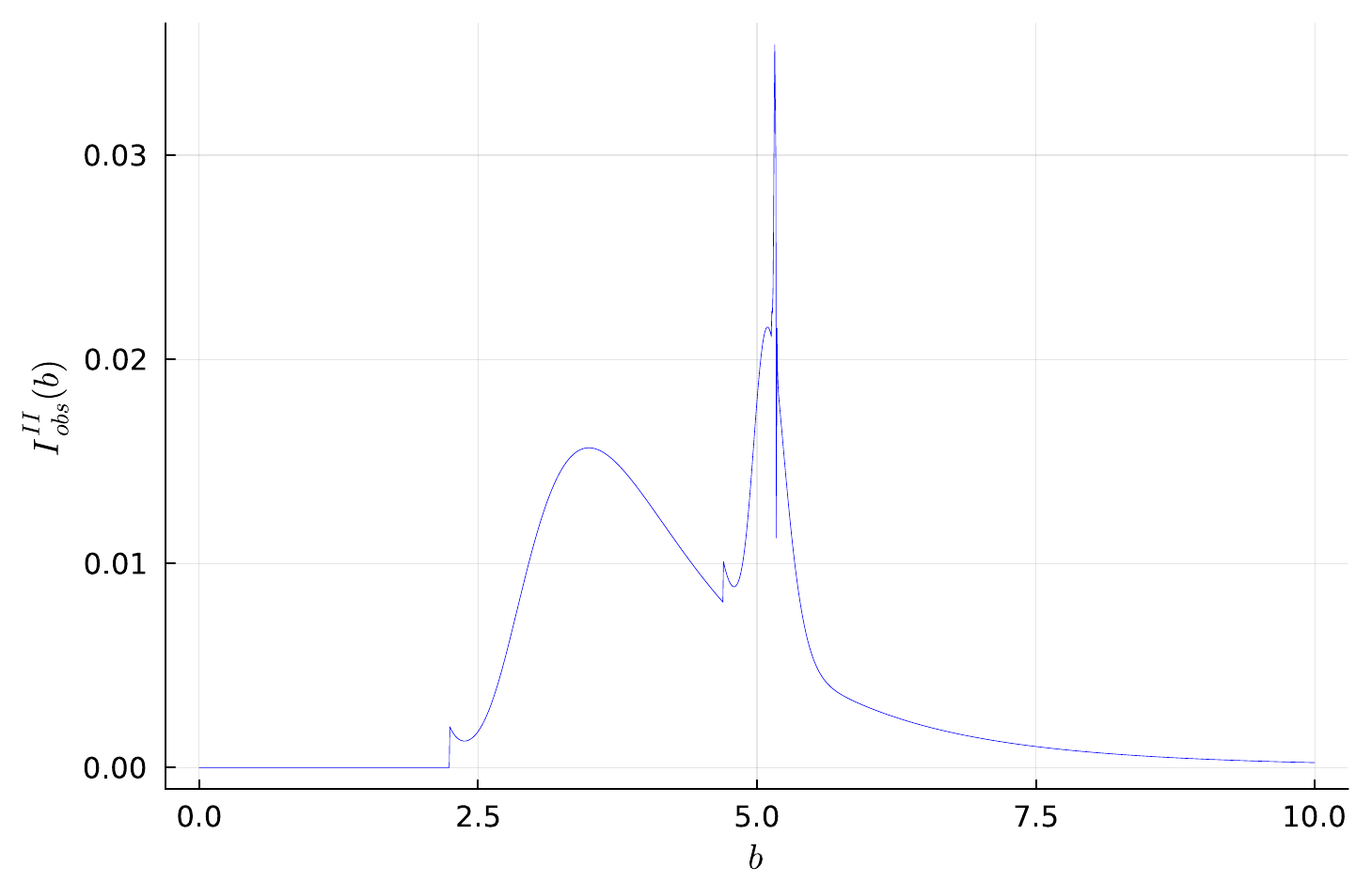}\,
    \includegraphics[width=0.32\linewidth]{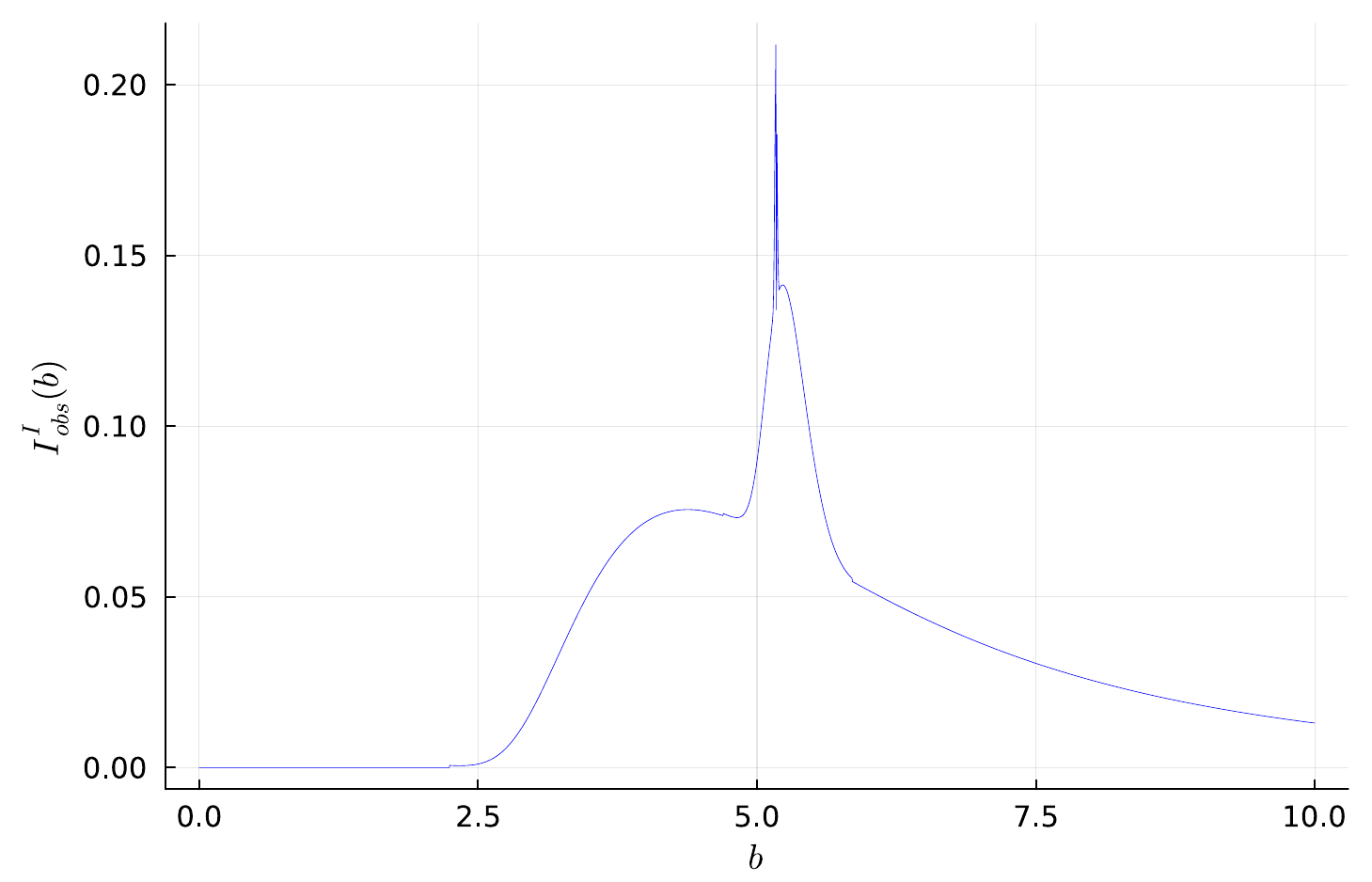}
    \caption{{Observed intensity as a function of impact parameter for the three emission profiles of Table~\ref{tab:GLM}. The top row corresponds to the two-horizon black hole with $\mathcal M=1$ and $\alpha=0.5$, whereas the bottom row corresponds to the static extremal configuration $\alpha=\alpha_E\simeq0.34856$. From left to right we show GLM1, GLM2, and GLM3. In both rows the same emissivity parameters are used in common ADM-mass units, with $x_0=2\mathcal M$, so that differences between the rows arise from the spacetime geometry rather than from a rescaling of the emission model.}}
    \label{fig: profile}
\end{figure*}

\begin{figure*}
    \centering
    \includegraphics[width=0.32\linewidth]{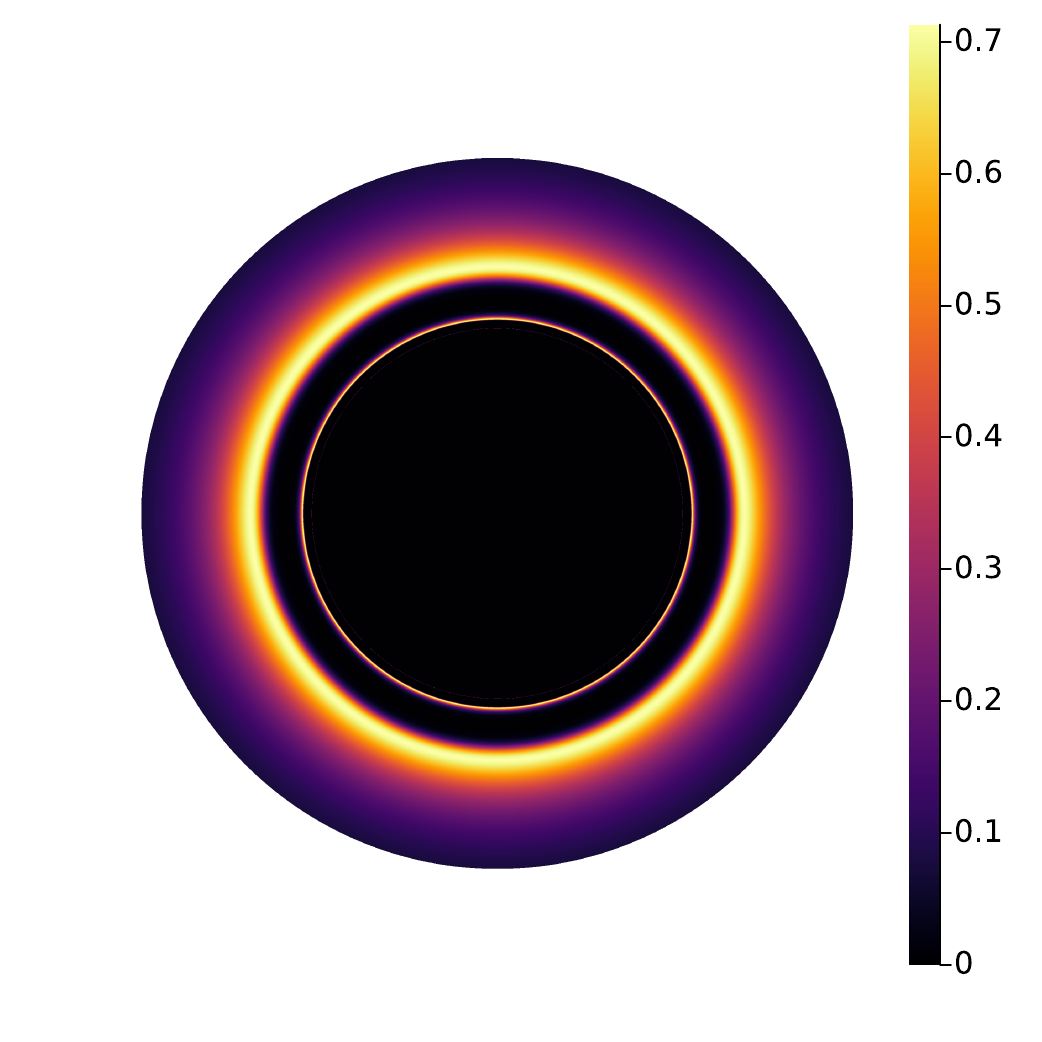}\,
    \includegraphics[width=0.32\linewidth]{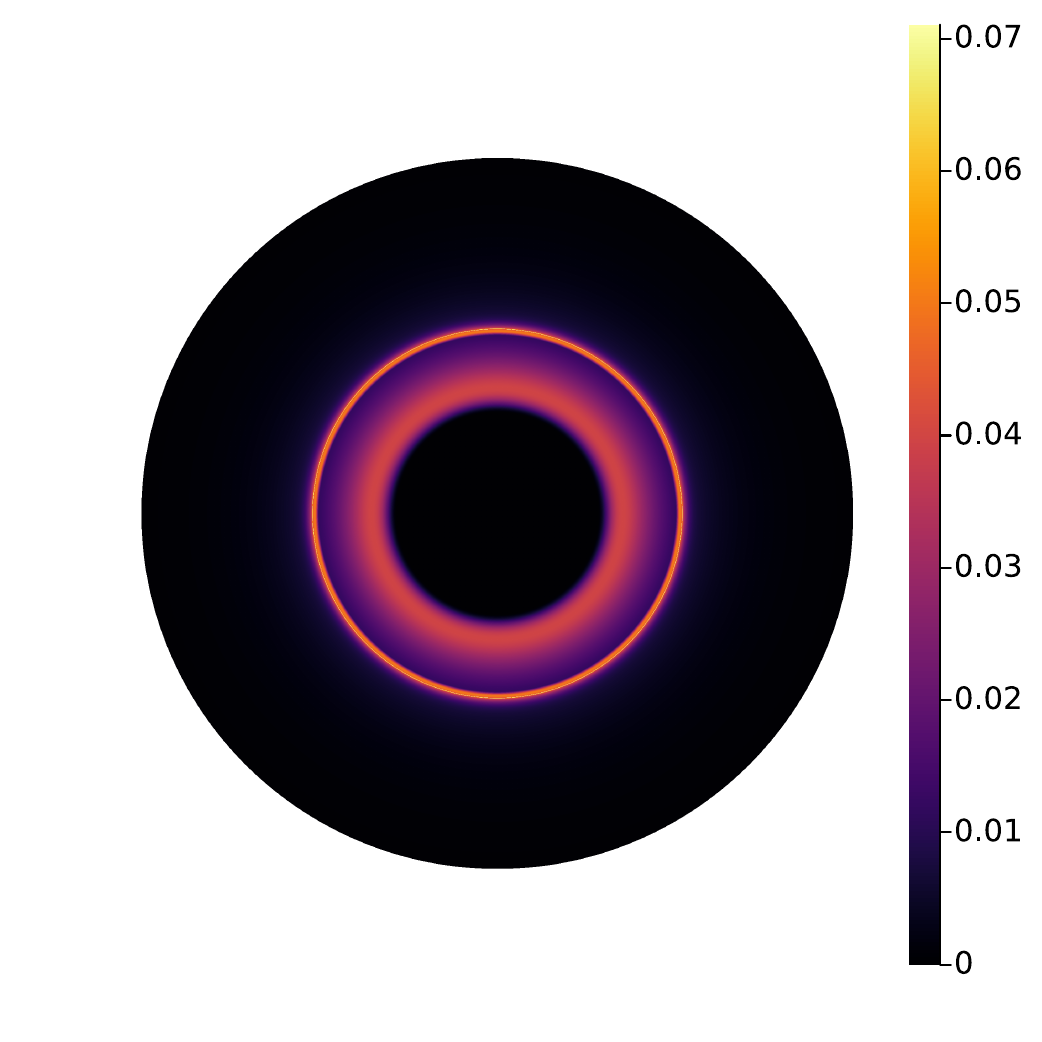}\,
    \includegraphics[width=0.32\linewidth]{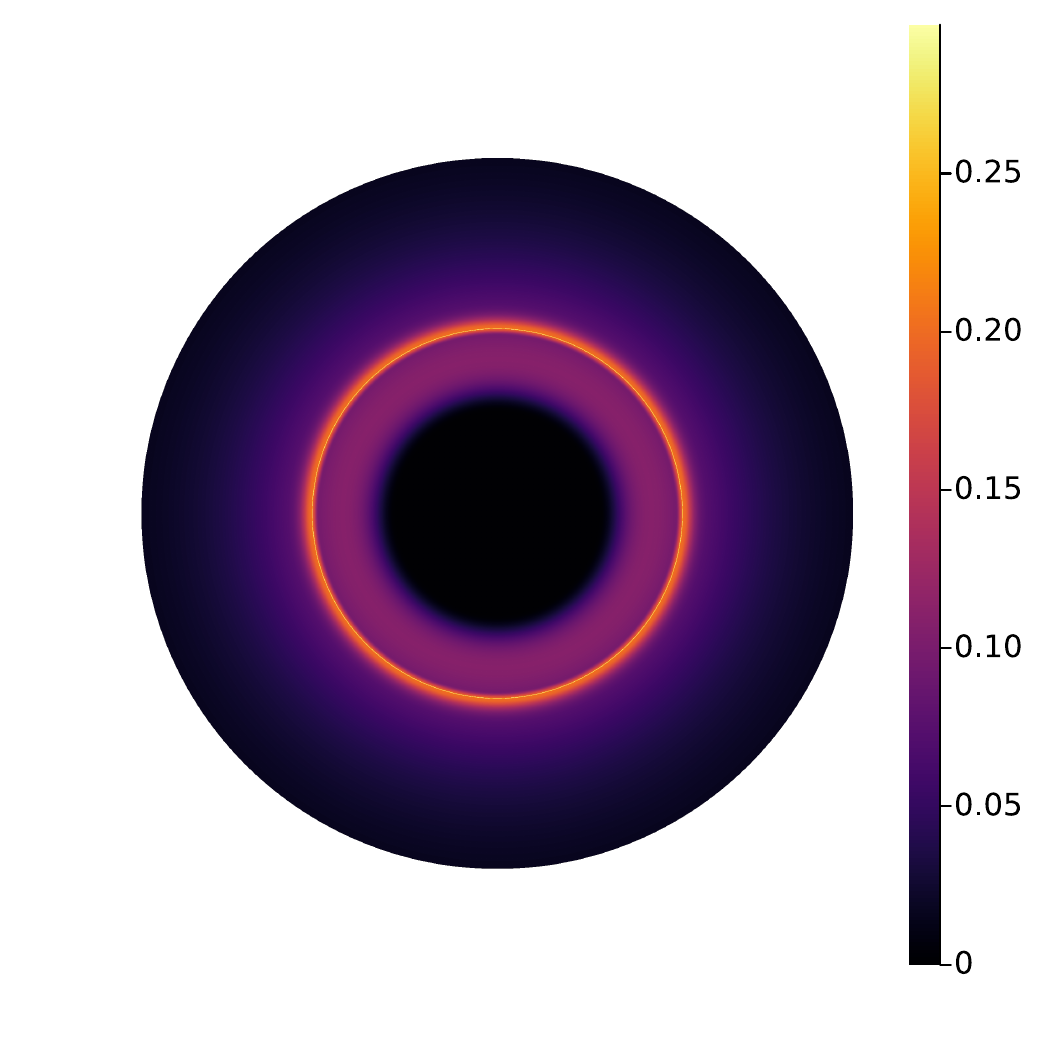}\,
    \includegraphics[width=0.32\linewidth]{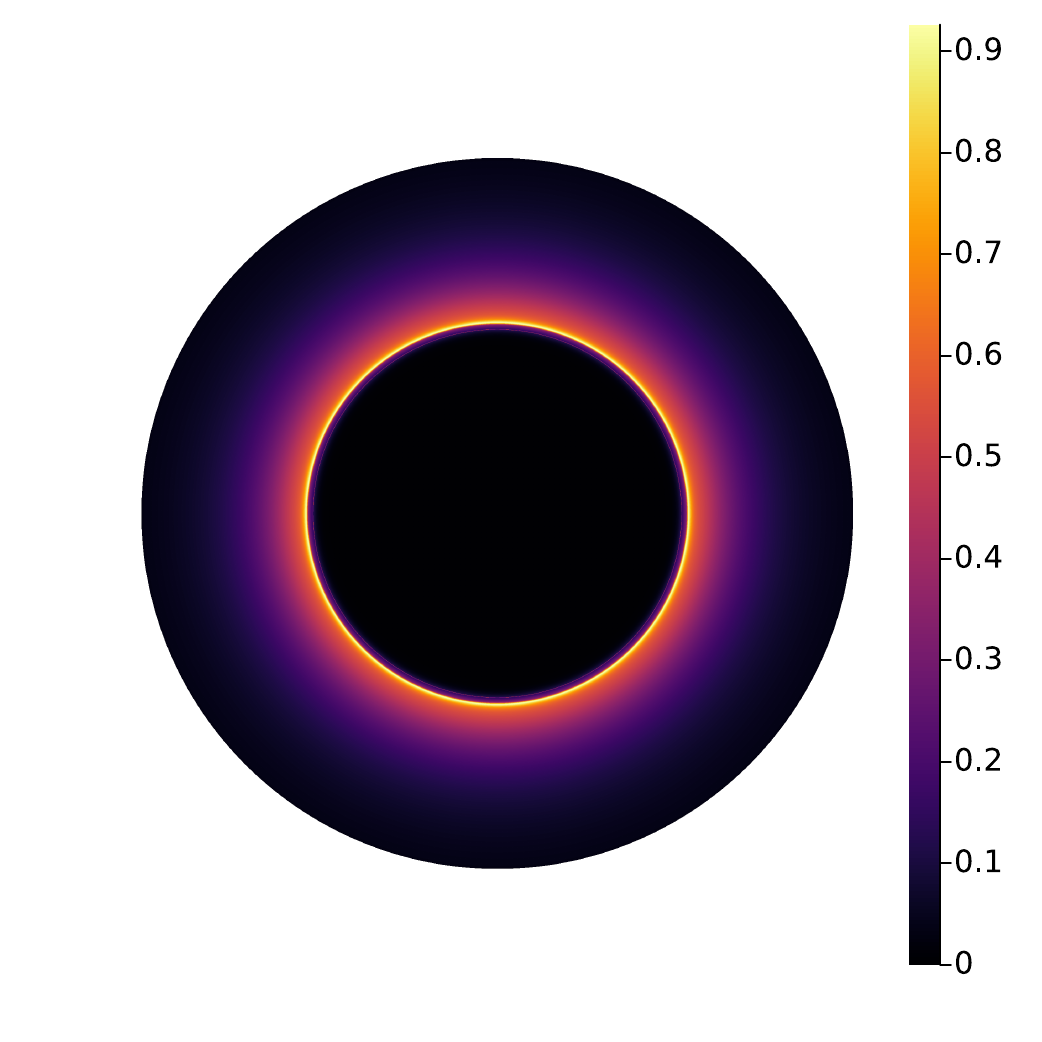}\,    \includegraphics[width=0.32\linewidth]{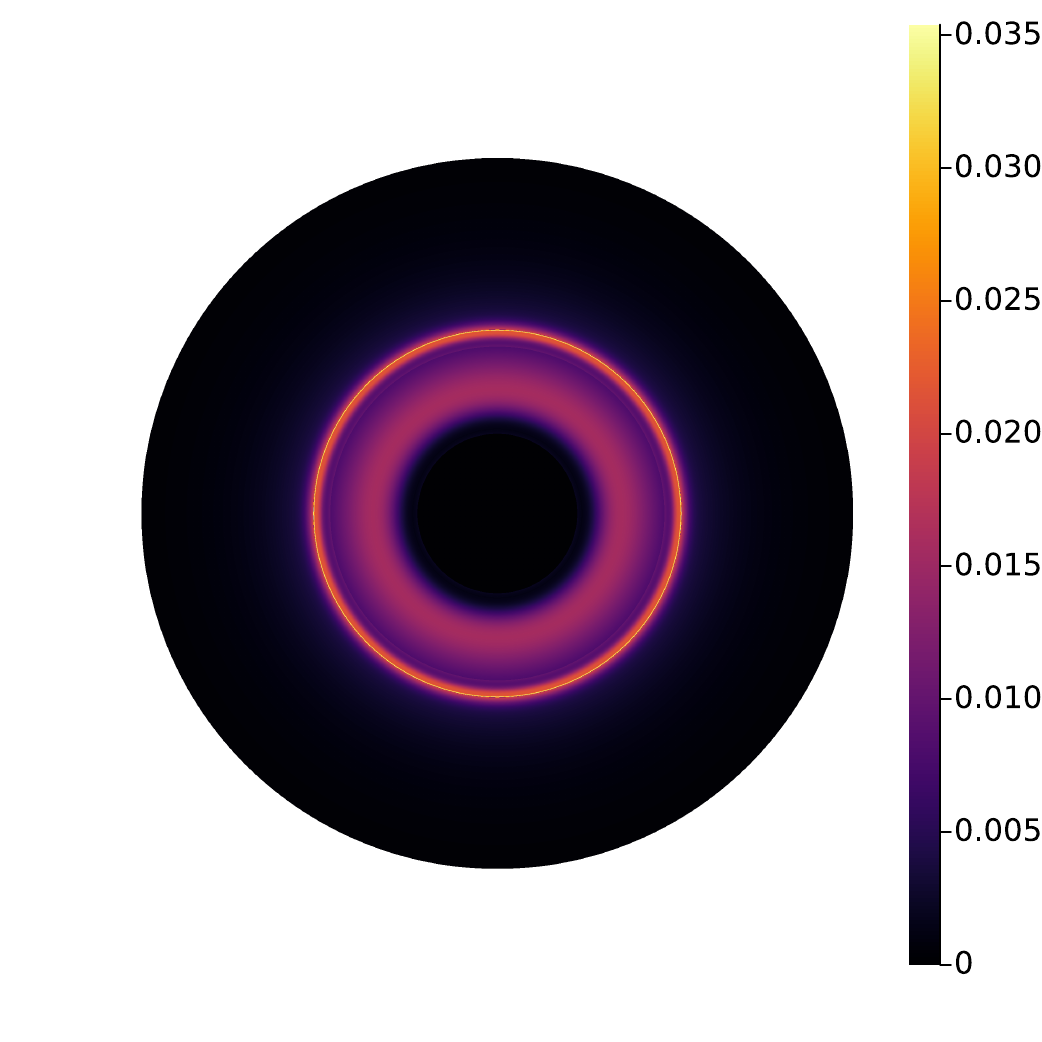}\,
    \includegraphics[width=0.32\linewidth]{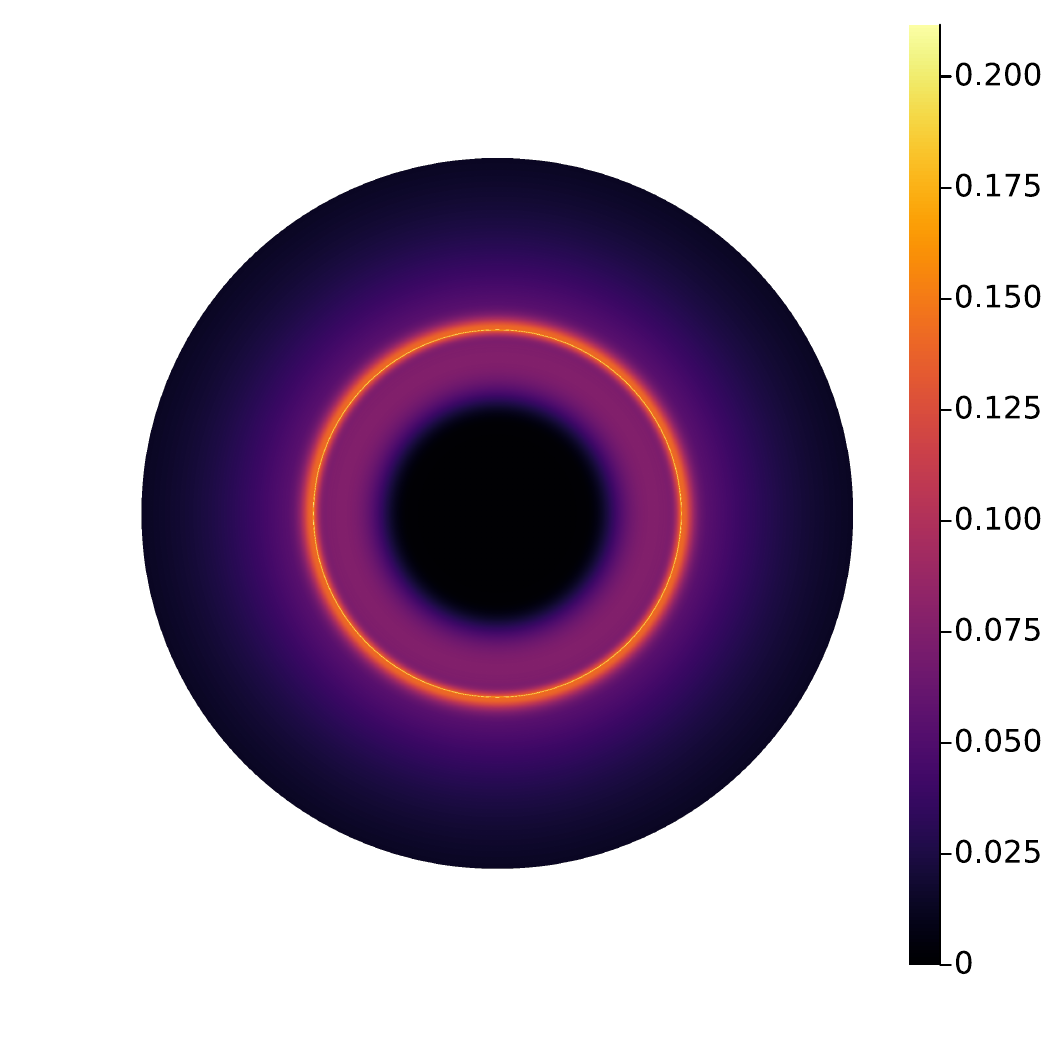}
    \caption{{Face-on ray-traced images of the static geometry for the three emission profiles of Table~\ref{tab:GLM}. The top row shows the two-horizon black hole with $\mathcal M=1$ and $\alpha=0.5$, and the bottom row the extremal configuration $\alpha=\alpha_E\simeq0.34856$; GLM1, GLM2, and GLM3 are shown from left to right. The emissivity law is held fixed in common ADM-mass units, $x_0=2\mathcal M$, in both rows. The central critical curve remains circular by spherical symmetry; changes in the surrounding brightness distribution reflect the transfer function and the adopted emissivity profile.}}
    \label{fig: images}
\end{figure*}
\begin{figure*}
    \centering
    \includegraphics[width=0.32\linewidth]{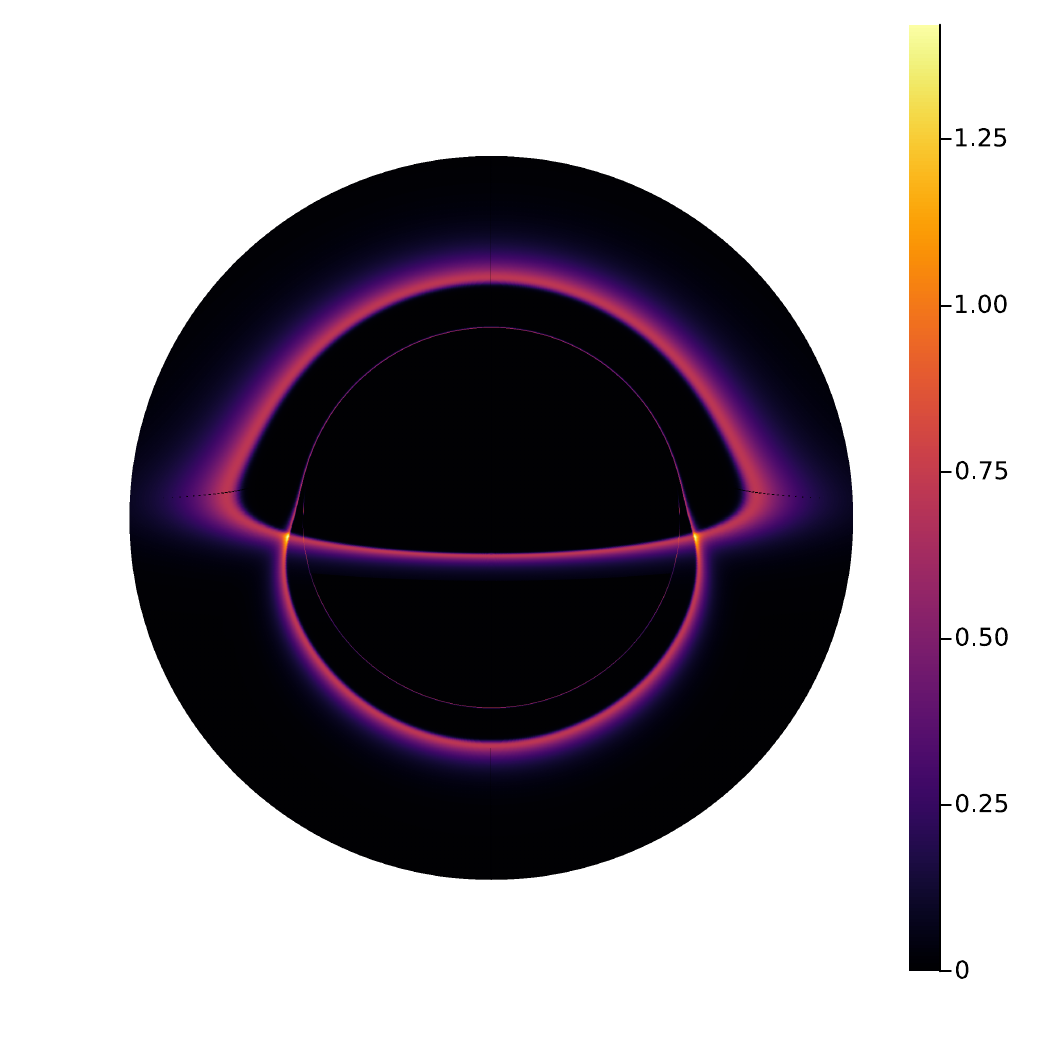}
    \includegraphics[width=0.32\linewidth]{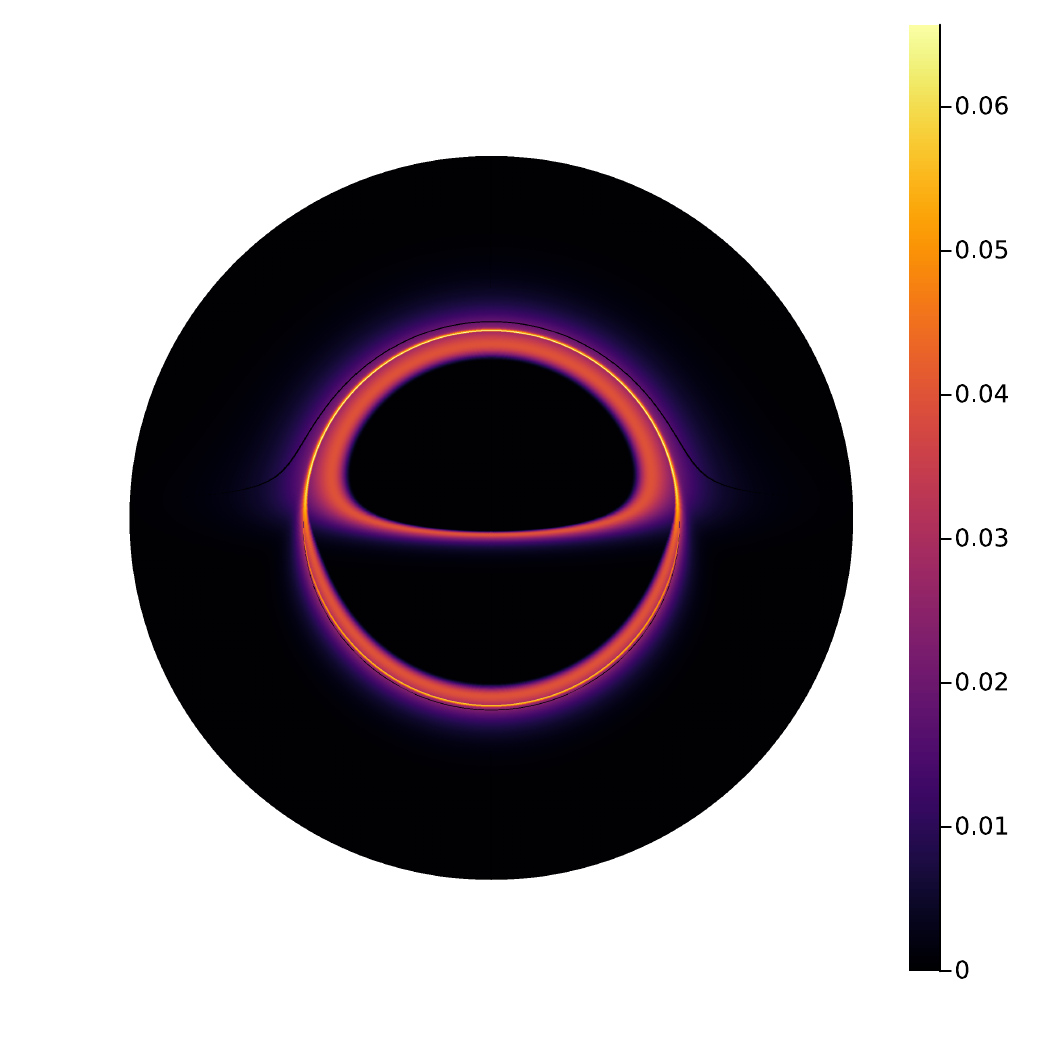}
    \includegraphics[width=0.32\linewidth]{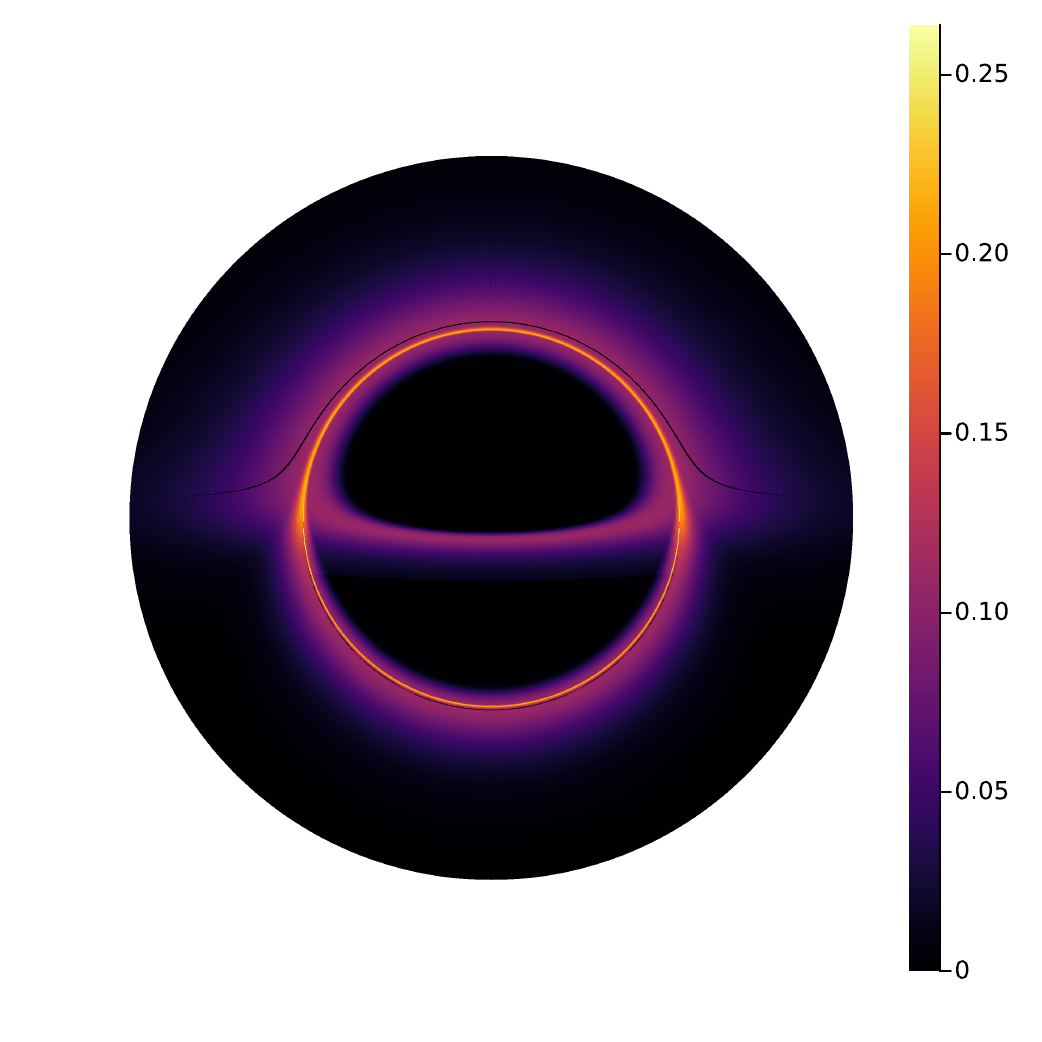}
    \includegraphics[width=0.32\linewidth]{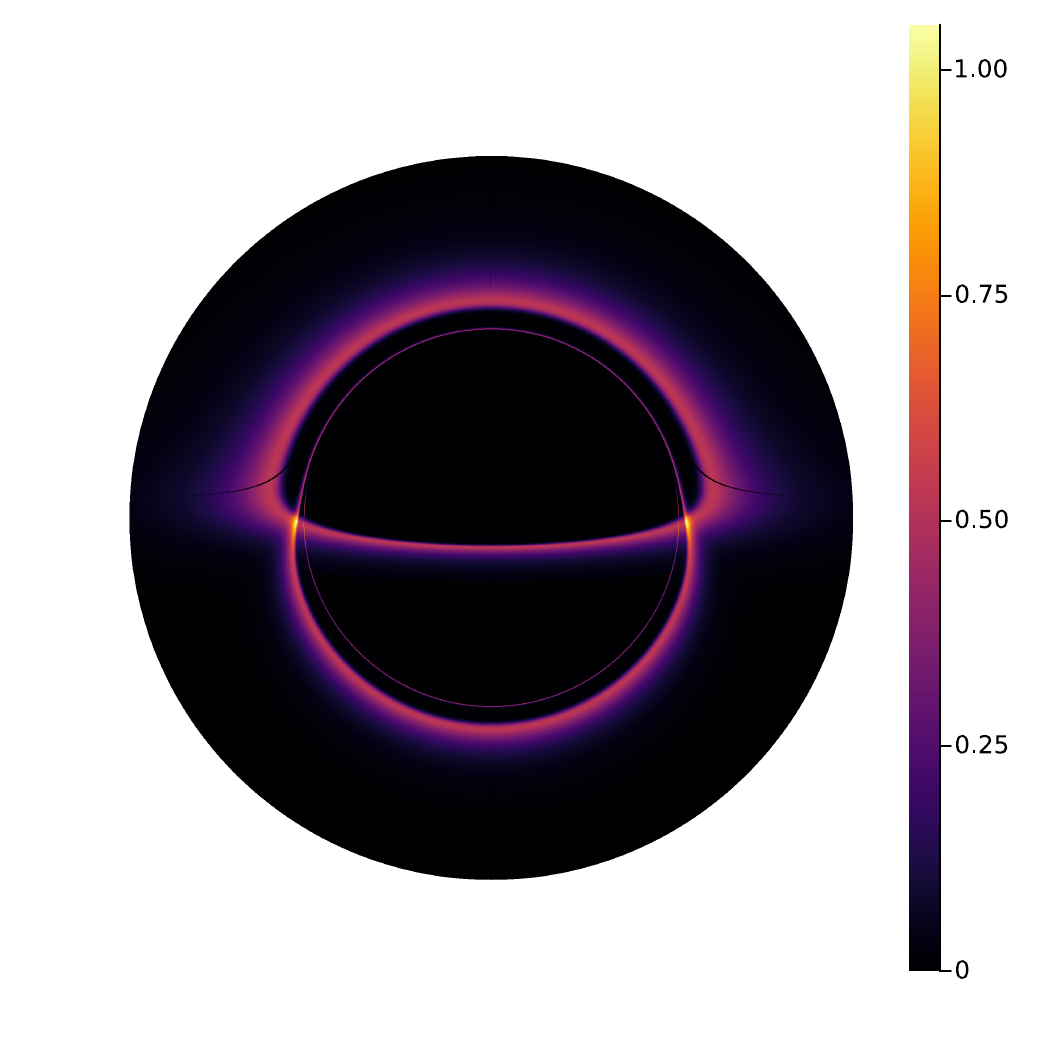}    \includegraphics[width=0.32\linewidth]{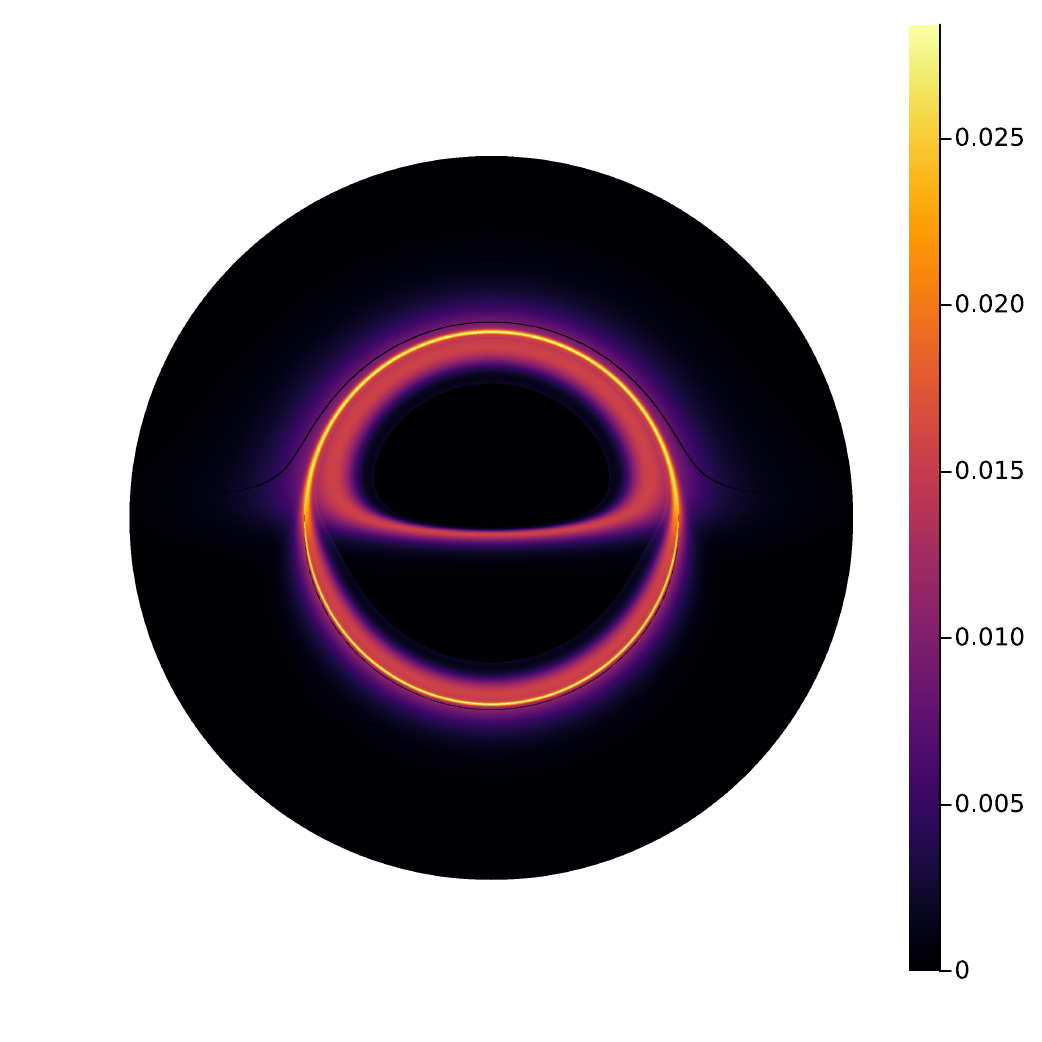}
    \includegraphics[width=0.32\linewidth]{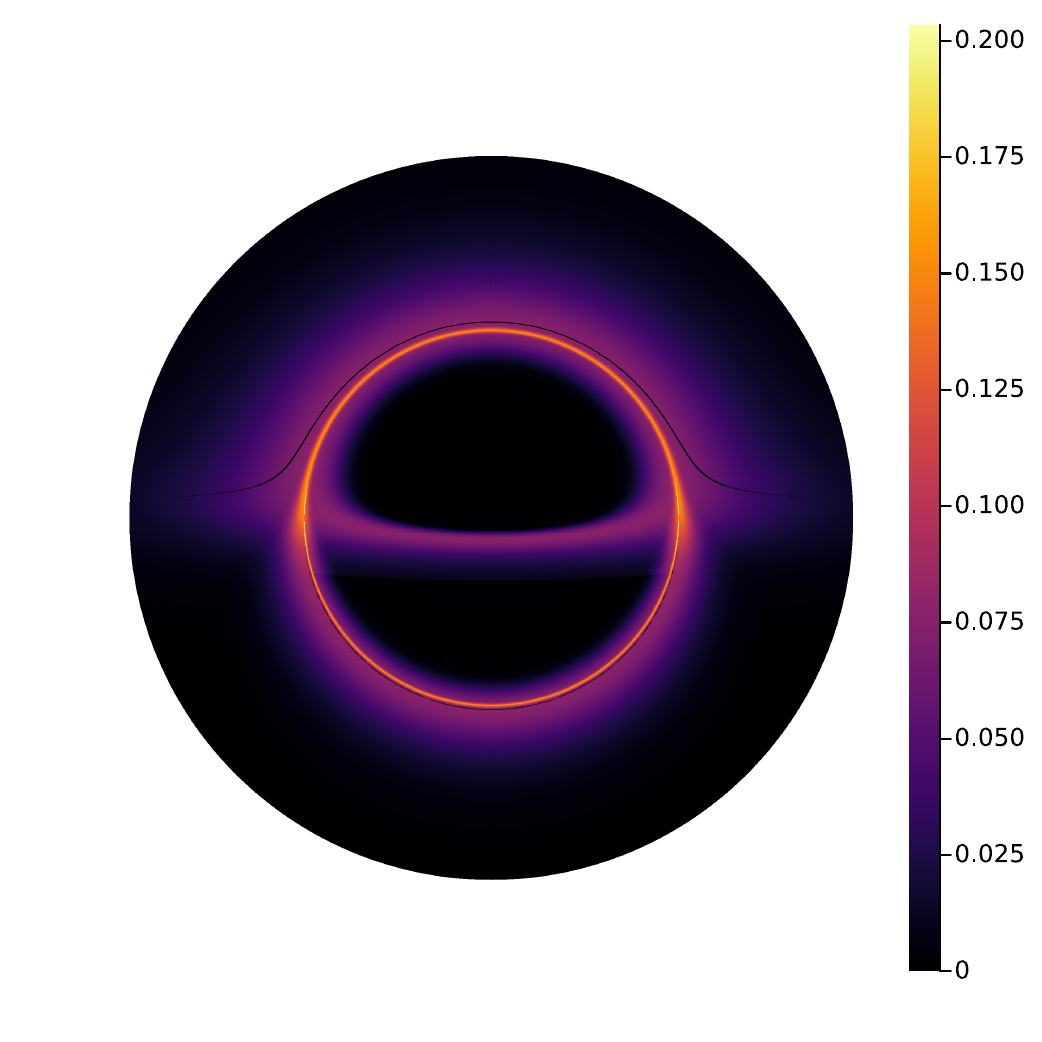}
    \caption{{Ray-traced images of the same static configurations as in Fig.~\ref{fig: images}, now viewed at the illustrative inclination $\iota=87^\circ$. The top row corresponds to $\mathcal M=1$, $\alpha=0.5$ and the bottom row to $\alpha=\alpha_E\simeq0.34856$; GLM1, GLM2, and GLM3 are shown from left to right, with the same fixed emissivity parameters and $x_0=2\mathcal M$. Because the spacetime and emitter prescription used here are static, the left--right brightness structure is produced by lensing and multiple disk crossings and must not be interpreted as Doppler boosting. The geometric critical curve itself remains circular at every inclination.}}
    \label{fig: images_inc}
\end{figure*}

\section{Observational appearance}\label{sec5}
\subsection{Spherically symmetric geometry}
For the static, spherically symmetric metric \eqref{yweakxx2}, the motion of photons is governed by the effective potential

\begin{equation}
V_{\rm eff}(r) = \frac{e^{\nu(r)}}{r^{2}},
\end{equation}
{where $e^{\nu}=e^{-\lambda}$ follows from the Schwarzschild-gauge closure condition imposed in Eq.~\eqref{yconstxx}.} The photon sphere radius $r_{\rm ph}$ is determined by the simultaneous conditions
\begin{equation}
V_{\rm eff}(r_{\rm ph}) = b^{-2}, \qquad V_{\rm eff}'(r_{\rm ph}) = 0,
\end{equation}
where $b$ is the impact parameter. The critical impact parameter $b_{\rm c}=r_{\rm ph}/\sqrt{e^{\nu(r_{\rm ph})}}$ sets the boundary between captured and escaping photon trajectories, defining the outer edge of the BH shadow.

{The incomplete-gamma representation provides an analytic explanation for the small optical deviation. At any radius $r$ the fractional mass deficit is
\begin{equation}
\delta_m(r)\equiv\frac{\mathcal M-\tilde m(r)}{\mathcal M}=Q(5,r/\ell).
\end{equation}
On the black-hole branch the photon sphere remains close to $r_{\rm ph}\simeq3\mathcal M$, so that
\begin{equation}
\delta_m(r_{\rm ph})\simeq Q(5,36\alpha)
=e^{-36\alpha}\sum_{k=0}^{4}\frac{(36\alpha)^k}{k!}.
\label{opticalscreen}
\end{equation}
Because the Schwarzschild photon sphere is a stationary point of $r^2/F(r)$, the displacement of $r_{\rm ph}$ does not contribute to the critical impact parameter at first order. One consequently obtains
\begin{equation}
\frac{\delta b_c}{b_c}\simeq-Q(5,36\alpha).
\label{bcshift}
\end{equation}
For $\alpha=0.5$, Eq.~\eqref{bcshift} gives $-8.42\times10^{-5}$, in excellent agreement with the full numerical value $-8.45\times10^{-5}$. Even close to the static extremal branch, $\alpha\simeq0.35$, the estimate remains at the sub-percent level. The near-degeneracy is therefore not accidental: it follows from exponential localization of the source well inside the photon region.}

The observational appearance of a compact object is shaped by the properties
of its surrounding accretion disk, whose emission is governed in general by
the absorptivity $\alpha_\nu$, specific intensity $I_\nu$, and emissivity
$j_\nu$, coupled through the radiative transfer equation \cite{Misner1973}.
In what follows we adopt a simplified accretion model that has been widely
employed in the literature Ref.~\cite{Luminet1979,Gralla2019,GLM2020}, and whose
key assumptions we summarize below.

\textit{Geometrically thin disk.} The disk is assumed to have negligible
vertical extent, so that emission is confined to the equatorial plane.
This choice is motivated by the observation that the photon-ring structure
and the central brightness depression produced by thin-disk models are
qualitatively more consistent with the corresponding features seen in
time-averaged GRMHD images than those produced by spherical accretion
models \cite{Gralla2019}. 

\textit{Optically thin disk.} We set the disk absorption to zero, so that
a given light ray may cross the equatorial plane multiple times, producing a
sequence of photon rings superimposed on the direct emission. 

\textit{Monochromatic rest-frame emission.}---The disk emits monochromatically
in its rest frame, with a specific intensity $\tilde{I}(r)$ that depends only
on the radial coordinate. We describe it by a suitable adaptation of the
Johnson Standard Unbound distribution~\cite{GLM2020,Vincent2022}
\begin{equation}
    \tilde{I}(r;\,\gamma,\mu,\sigma)
    =
    \frac{\displaystyle
    \exp\!\left\{-\tfrac{1}{2}\left[\gamma
    +\operatorname{arcsinh}\!\left(\tfrac{r-\mu}{\sigma}\right)\right]^{2}\right\}}
    {\sqrt{(r-\mu)^{2}+\sigma^{2}}},
    \label{eq:Johnson}
\end{equation}
previously employed to reproduce time-averaged GRMHD images of M87$^*$ and
Sgr~A$^*$~\cite{GLM2020,Vincent2022}. We consider three
{emission profiles whose parameters are listed in Table~\ref{tab:GLM}. Their functional forms follow the GLM family, but for the controlled comparisons performed here the radial scale is kept fixed in units of the common ADM mass rather than being re-anchored to an $\alpha$-dependent horizon or ISCO. This prevents changes in the source prescription from being confused with changes in the geometry.}

\begin{table}[H]
\centering
\caption{Parameters of the three GLM emission profiles used in this work.
{To avoid changing the emission model when $\alpha$ is varied, $x_0$ is fixed in units of the ADM mass rather than tied to the model-dependent horizon. In all comparisons below we use $x_0=2\mathcal M$ as a common reference scale.}}
\label{tab:GLM}
\begin{tabular}{lccc}
\hline\hline
Model & $\mu$ & $\sigma$ & $\gamma$ \\
\hline
GLM1 & $x_0$           & $1.5$  & $-1.5$ \\
GLM2 & $x_0$           & $0.5$  & $0$    \\
GLM3 & $17\,x_0/6$     & $0.25$ & $-2$   \\
\hline\hline
\end{tabular}
\end{table}

The observed intensity at the detector plane
is obtained by summing the redshifted contributions from each successive
crossing of the disk \cite{Luminet1979,Gralla2019}
\begin{equation}
    I_{\rm obs}(b)
    =
    \sum_{i=0}^{n}
    \,A^{2}(r)\,\tilde{I}(r)\Big|_{r\,=\,r_i(b)},
    \label{eq:Iobs}
\end{equation}
where the sum runs over the number of disk crossings, $r_i(b)$ is the
transfer function giving the equatorial radius at which a photon of impact parameter $b$ hits the disk on its $i$-th crossing, and $A^{2}(r)$ is the gravitational redshift factor.

{The resulting intensity profiles are shown in Fig.~\ref{fig: profile} for the two-horizon configuration $\alpha=0.5$ (top row) and the extremal configuration $\alpha=\alpha_E$ (bottom row). The three columns correspond to the GLM1--GLM3 emissivity prescriptions in Table~\ref{tab:GLM}. In the corrected comparison the emissivity parameters are identical in the two geometries and are expressed in common ADM-mass units. The direct component and the narrow features generated by higher-order disk crossings can therefore be compared without introducing an $\alpha$-dependent rescaling of the source. The location of the critical feature changes only weakly between the two rows, consistently with the exponentially small shift of $b_c$ derived in Eq.~\eqref{bcshift}. Differences in peak height or detailed profile shape should instead be interpreted as properties of the transfer function for the specified emissivity model, not as universal geometric observables.}

{Figure~\ref{fig: images} shows the corresponding face-on ray-traced maps. Spherical symmetry fixes the geometric critical curve to be circular, while the surrounding intensity distribution depends on the chosen GLM profile and on the sequence of disk crossings. Accordingly, these maps are used to illustrate the radiative appearance of the geometry rather than to define an additional shadow observable. The controlled comparison is between images generated with the same emission prescription at $\alpha=0.5$ and $\alpha=\alpha_E$; under this restriction, the small displacement of the critical scale is the robust geometric effect, whereas changes in brightness morphology remain model dependent.}

{Figure~\ref{fig: images_inc} presents the same static spacetime viewed at the illustrative high inclination $\iota=87^\circ$. This angle is used only to expose lensing morphology and is not intended to reproduce the preferred viewing geometry of either M87$^*$ or Sgr~A$^*$. Because the spacetime is static and the transfer prescription in Eq.~\eqref{eq:Iobs} contains gravitational redshift only, these images must not be interpreted in terms of approaching/receding Doppler boosting. The bright arcs instead arise from the lensing and multiple equatorial crossings of the far and near sides of the emitting disk.}

{For a spherically symmetric metric the critical curve itself is a circle of radius $b_c$ for every observer inclination. Any apparent noncircular structure in an inclined intensity map belongs to the lensed image of the emitting disk, not to a deformation of the shadow critical curve. This distinction is essential when separating geometry from astrophysical emission. Taken together, the static calculations support a much narrower statement than an EHT-specific claim: once a common emissivity profile is used, the critical scale is weakly sensitive to $\alpha$, consistently with the analytic screening law in Eq.~\eqref{bcshift}.}

\begin{figure*}
    \centering
    \includegraphics[width=0.45\linewidth]{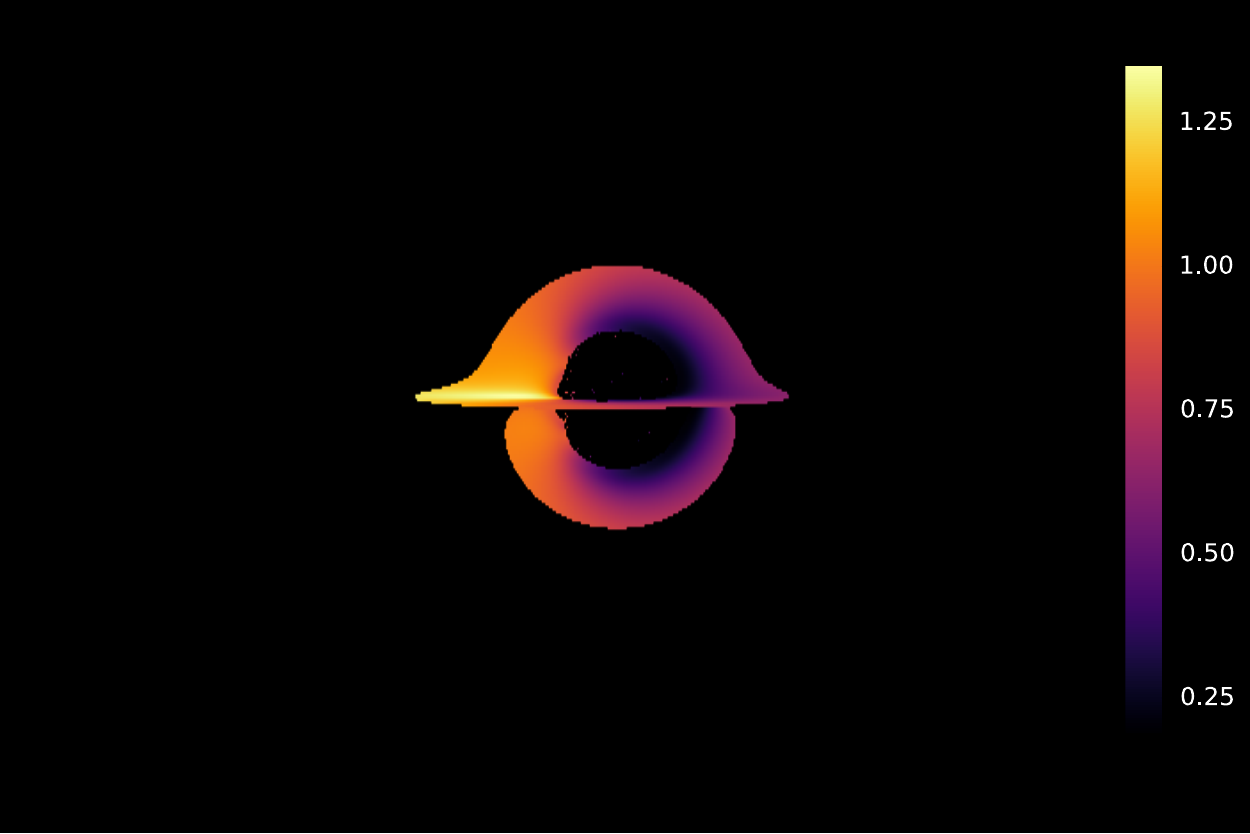}
    \includegraphics[width=0.45\linewidth]{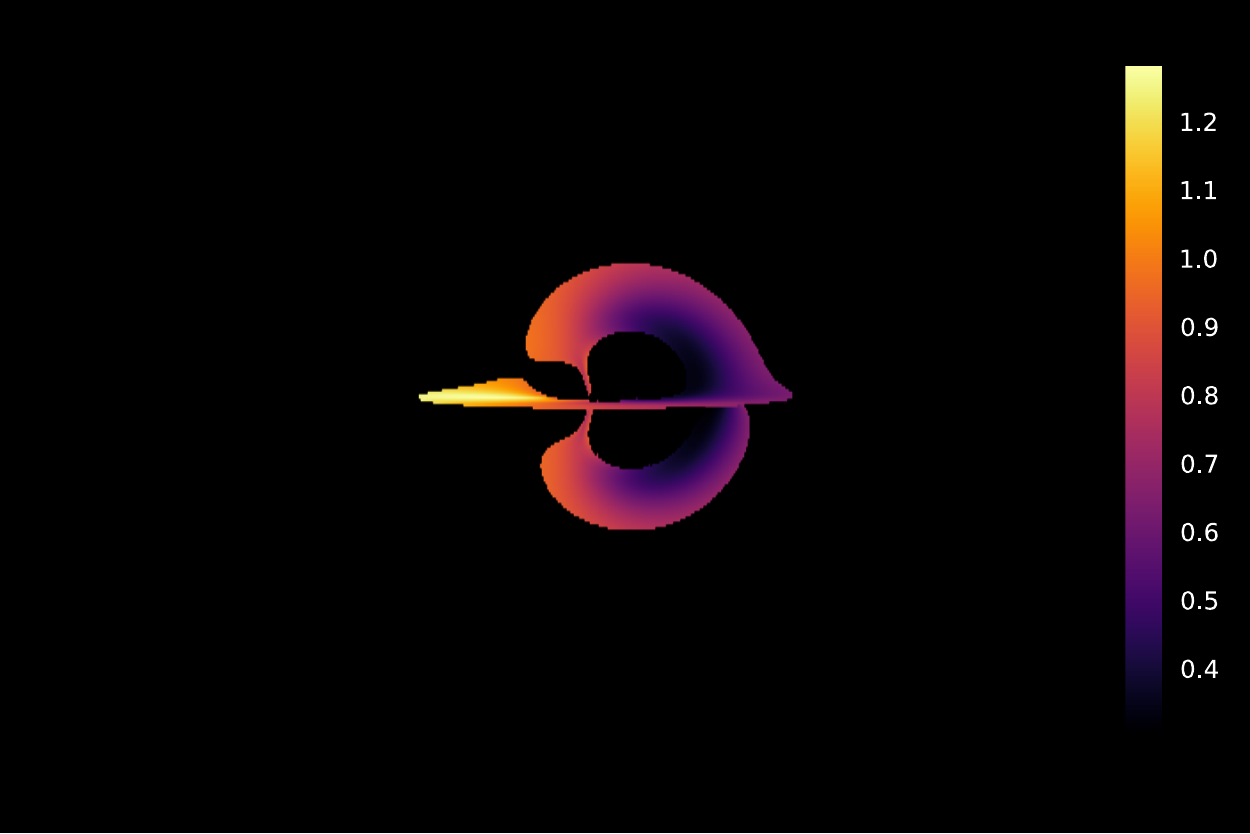}

    \caption{{Ray-traced appearance of the rotating G\"{u}rses--G\"{u}rsey geometry for an observer at $\iota=87^\circ$. The left panel uses $\{\mathcal M,a,\alpha\}=\{1,0.5,0.5\}$ and the right panel $\{1,0.95,1\}$; both parameter sets lie on the black-hole branch. The same disk-emission prescription and normalization must be used in the two panels. The high inclination is chosen to expose frame-dragging and lensing morphology and is not intended as an EHT-specific viewing angle. For an orbiting emitter, the intensity asymmetry combines gravitational redshift, Doppler boosting, and the lensing transfer function.}}
    \label{fig: images_rot}
\end{figure*}

\subsection{Axially symmetric geometry}

For the rotating case, we perform a fully numerical ray-tracing computation using the \textit{Gradus.jl} package in Julia \cite{Baker2026}, which integrates the geodesic equations in the G\"{u}rses--G\"{u}rsey metric \ref{kerrex} with the mass function \ref{mass1}. {For the illustrative maps shown here the observer is placed at $\iota=87^\circ$ with respect to the spin axis. This high-inclination choice is a morphology test and is not identified with the preferred EHT viewing angles. {Indeed, the EHT modeling of M87$^*$ infers an inclination of only $\sim17^\circ$ from the large-scale jet~\cite{Walker2018}, and the EHT model comparison for Sgr~A$^*$ disfavors inclinations $i\gtrsim70^\circ$~\cite{EHTSgrAModels2022}; an edge-on choice such as $\iota=87^\circ$ is therefore used here purely as a morphology stress-test, not as a representative viewing angle for either source.}}

{Figure~\ref{fig: images_rot} is defined using two parameter sets that lie on the black-hole branch: $\{\mathcal M,a,\alpha\}=\{1,0.5,0.5\}$ and $\{1,0.95,1\}$. The latter replaces the previously used horizonless pair $(a,\alpha)=(0.95,0.7)$. The comparison is intended to illustrate how increasing spin modifies the lensed disk morphology while remaining within the admissible black-hole domain. Since the rotating calculation employs an orbiting emitter, the observed asymmetry may contain both Doppler and gravitational-redshift effects in addition to lensing. The two panels should be generated with the same emission prescription so that their morphological differences are not contaminated by a change of source model.}

\section{Conclusions}\label{sec6}

{We have constructed and analyzed an asymptotically flat geometry whose static branch replaces the Schwarzschild singularity by a Minkowskian core. The regularity condition fixes $\ell/\mathcal M=(12\alpha)^{-1}$ and allows the exact mass profile to be written as $\tilde m=\mathcal M P(5,r/\ell)$. This representation makes the physics of the solution considerably more transparent: $\alpha$ controls the localization scale of the core rather than acting as a conventional perturbative deformation strength.}

{A central result concerns the effective matter sector. The relation ${\cal P}_r=-{\cal E}$ implies ${\cal E}+{\cal P}_\theta=-r{\cal E}'/2$. Since a positive Minkowskian-core density vanishes at the origin and must initially increase, the tangential NEC is necessarily violated in an inner neighborhood. For the explicit profile used here the violating region is $0<r<2\ell$. The DEC is even more restrictive, holding only for $2\leq r/\ell\leq6$. Thus the model should not be advertised as globally satisfying the standard energy conditions; rather, it provides a concrete realization of the tension between a vanishing-density Minkowskian core and the NEC in this static closure class.}

{The horizon structure is governed by a conventional fold bifurcation. On the static black-hole branch a finite-radius minimum of $F$ lies below zero and produces an inner and an outer horizon; at $\alpha_E\simeq0.34856$ they merge. The existence of the inner horizon is therefore not, by itself, a new causal mechanism. What differs from de Sitter-core models is the local interior geometry: the density and curvature vanish at the center instead of approaching an effective cosmological-constant core. The stability of the resulting Cauchy horizon remains an open dynamical problem.}

{The second main result is the analytic optical-screening relation. The mass deficit outside radius $r$ is exactly $Q(5,r/\ell)$, and at the photon sphere this gives the exponentially small quantity $Q(5,36\alpha)$. To first order, the critical-impact-parameter deviation obeys $\delta b_c/b_c\simeq-Q(5,36\alpha)$, accurately reproducing the numerical result. This explains why a strongly modified core can coexist with an almost Schwarzschild exterior optical scale: by the time null rays reach the photon region, the mass function has essentially saturated to its ADM value.}

{For the rotating G\"urses--G\"ursey extension we have also clarified two limitations. First, extremality must be obtained from $\Delta=\partial_r\Delta=0$; the previously used pair $(a,\alpha)=(0.95,0.7)$ is horizonless, while $a=0.95$ requires $\alpha_{\rm ext}\simeq0.71192$. Second, finiteness of curvature invariants on $r\geq0$ does not establish a globally regular analytic extension through the Kerr disk. We therefore restrict our statement to curvature regularity on the physical $r\geq0$ domain and leave geodesic completeness and possible extensions as separate questions. {This is consistent with the general obstruction to a smooth analytic extension through the ring singularity established in Refs.~\cite{ZhouModesto2023,TorresFayos2017} for mass functions without definite parity in $r$, and the off-equatorial enlargement of the energy-condition violation expected on general grounds from Ref.~\cite{NevesSaa2014}.}}

{From the standpoint of exact solutions, the significance of the construction is therefore not that Minkowskian cores or Kerr-like regularizations are new by themselves. Rather, this particular incomplete-gamma profile gives a single analytic framework in which the core energy-condition obstruction, the fold structure of the horizons, and the exterior optical screening can be traced to the same localized mass distribution. This is the geometric relation that distinguishes the present analysis from simply cataloguing another regular metric.}

{Finally, the imaging analysis must distinguish geometric critical curves from model-dependent intensity morphology. In the static spacetime the critical curve remains circular at every inclination, and no Doppler asymmetry arises from a static-emitter transfer model. Comparisons between different values of $\alpha$ are meaningful only when the emissivity law is held fixed in common ADM-mass units. With these controls, the robust conclusion is not that every horizon-scale image is universally degenerate, but that the exterior critical scale is exponentially insensitive to the Minkowskian core. A direct, controlled comparison with a de Sitter-core geometry at the same ADM mass and with an identical emission prescription is the natural next test of how much information about the core can survive in strong-lensing observables.}

\section*{Acknowledgments}
Y. Gómez-Leyton acknowledges to ANID subvención en la academia convocatoria año 2025 85250186.
The work of Kazuharu Bamba was supported in part by the JSPS KAKENHI Grants No. 24KF0100 and No. 25KF0176, and a grant-in-aid of academic research of the Yamaguchi Scholarship Foundation.

\bibliography{references.bib}
\bibliographystyle{apsrev4-1.bst}

\end{document}